\documentclass[10pt,a4paper,notitlepage]{article}
\usepackage[utf8]{inputenc}
\usepackage{amsmath,amsfonts,amssymb}
\usepackage{graphicx}
\usepackage{hyperref}
\usepackage[english]{babel}
\usepackage{epstopdf}
\usepackage{multirow}
\usepackage{tabularx}
\usepackage{floatrow}
\usepackage{subcaption}

\hypersetup{
	colorlinks=true,         
	linkcolor=blue,          
	citecolor=Violet,        
	urlcolor=Violet            
}
\usepackage[usenames,dvipsnames]{color}

\begin{document}

	\title{ 	
		\Large \bf{Visible and Invisible Pseudoscalar Meson Decays from Anomaly Sum Rules} }
	
	\author{Xurong Chen,$^1$~~
Sergey~Khlebtsov,$^2$~~
			Armen~Oganesian,$^{2,3}$~~
		Oleg~Teryaev$^{1,2,3}$
		\footnote{Electronic addresses: 
\href{mailto:xchen@impcas.ac.cn,}{ xchen@impcas.ac.cn},
			\href{mailto:khlebtsov@itep.ru,}{khlebtsov@itep.ru},
			 \href{mailto:armen@itep.ru,}{armen@itep.ru,} \href{mailto:teryaev@theor.jinr.ru}{teryaev@theor.jinr.ru}.}
		\vspace{12pt} \\
\it \small  $^1$Institute of Modern Physics, Chinese Academy of Sciences, Lanzhou 730000, China\\
		\it \small  $^2$
National Research Center "Kurchatov Institute", 123182,  Moscow, Russia\\
		\it \small $^3$ Bogoliubov Laboratory of Theoretical Physics,\\
		\it \small Joint Institute for Nuclear Research, 141980, Dubna, Russia}

	\date{}
	\maketitle

	\begin{abstract}
The decays of pseudoscalar mesons to real and virtual photons as well as neutrino-antineutrino pairs are considered in the framework of the dispersive method based on Anomaly Sum Rules. The contribution of singlet channel involving the new non-perturbative gluon form factor of virtual photon  $B(q^2)$ is systematically taken into account. The detailed analysis of its dependence on photon virtuality $q^2$ relying on the available data for meson transition fomfactors is performed. It is shown that B has quite  a nontrivial structure at $q^2 \sim 1 GeV^2$ which may be a signal of the existence of pseudoscalar glueball with a mass about 1.5-2 GeV. The calculation of the decay to $\nu \bar \nu$ pairs leads to the compatibility with the result of Arnellos, Marciano and Parsa of 1982, when pion decay is considered neglecting the mixing effects. The account for these effects results, however, in the enhancement of pion branching ratio by a factor of 3, while that for eta decay is larger by several orders of magnitude. It is stressed, that dependence on  the pair invariant mass is entirely defined by QCD and coincides with that of the meson transition form factor. The role of obtained results for the physics at HHaS detector at HIAF is discussed.  
	\end{abstract}
	\section{ Introduction} 
The main aim of this work is the precise theoretical study of various decays of pseudoscalar mesons, paying  the special attention to the semi-visible mode $\pi^0,\eta \rightarrow \nu \bar{\nu} \gamma$, being the potential background for decay to dark photons.  The first theoretical description of these latter processes was developed in 1982 by Arnellos, Marciano and Parsa \cite{Arnellos:1981bk}. We would like to revisit these processes from the modern perspective. It appears that simultaneously we will get the important knowledge on the Dalitz decays
to charged lepton  pairs and related gluon form factor of virtual photons.  

The crucial common point for calculation of visible and semi-visible decay modes is to have knowledge of Transition Form Factors(TFF) $\pi^0,\eta \rightarrow Z^0 \gamma$. In this paper it will be proved that they can be expressed in terms of $\pi^0,\eta \rightarrow \gamma^* \gamma$ TFFs, which was theoretically investigated and obtained by the dispersive method   \cite{Khlebtsov:2018roy, Khlebtsov:2020rte} based on Anomaly Sum Rules technique. Here we develop this method providing more involved parameterization of form factor $B(q^2)$. The advantage of such approach is the possibility of  TFF description   in the whole range of  photon virtualities. 

The paper is organised as follows. In Section 2 we describe main aspects of the  anomaly sum approach and suggest the new parametrization of the gluon form factor of virtual photon and  related duality intervals, required for more accurate data description. In Section 3 we discuss the specifics of TFFs for the case $\pi^0,\eta \rightarrow Z^0 \gamma$  and derive the accurate numerical results for the widths of $\pi^0,\eta \rightarrow \nu \bar{\nu} \gamma$ decays.   Section 4 is dedicated to discussion and conclusions.   

\section{Anomaly Sum Rules}
	It is well known that the axial anomaly \cite{Adler:1969gk, Bell:1969ts}
results in the non-conservation of the axial current. In particular, the singlet axial current  $J_{\mu5}^{(0)}=(1/\sqrt{3})\sum_i {\bar{\psi_i}\gamma_{\mu}\gamma_5 \psi_i}$ gets contributions from both electromagnetic and strong (non-Abelian) anomalies, while isovector ($a=3$) and octet ($a=8$) components of the octet of axial currents  $J_{\mu5}^{(a)}=(1/\sqrt{2})\sum_i {\bar{\psi_i}\gamma_{\mu}\gamma_5\lambda^a \psi_i}$  acquire only the electromagnetic anomaly:
	\begin{equation}\label{an-0}
	\partial^\mu J_{\mu 5}^{(0)} =\frac{2i}{\sqrt 3}\sum_i{m_i \bar{\psi_i} \gamma_5 \psi_i}+\frac{e^2}{8\pi^2}C^{(0)}N_c  F\widetilde{F} + \frac{n_f\alpha_s}{4\pi\sqrt{3}}  G\widetilde{G},
	\end{equation}
	
	\begin{equation}\label{an-38}
	\partial^\mu J_{\mu 5}^{(a)} =\frac{2i}{\sqrt2}\sum_i{m_i \bar{\psi_i} \gamma_5 \lambda^a \psi_i} 
	+\frac{e^2}{8\pi^2}C^{(a)}N_c  F\widetilde{F}, \; a=3,8.
	\end{equation}
	Here $F$ and $G$ are electromagnetic and gluon field strength tensors respectively,  $\widetilde{F}^{\mu\nu}=\frac{1}{2}\epsilon^{\mu\nu\rho\sigma}F_{\rho\sigma}$ and $\widetilde{G}^{\mu\nu,t}=\frac{1}{2}\epsilon^{\mu\nu\rho\sigma}G_{\rho\sigma}^{t}$ are their duals, $N_c=3$ is a number of colors, $n_f=3$ is the number of flavors, $\alpha_s$ is a strong coupling constant, $C^{(a)}$ are the charge factors ($e_i$ are quark charges in units of the electron charge $e$):
\begin{align}\label{ch_fctrs_2gamma}
C^{(3)}=&\frac{1}{\sqrt 2} (e_u^2-e_d^2)=\frac{1}{3\sqrt 2},\\ C^{(8)}=&\frac{1}{\sqrt 6} (e_u^2+e_d^2-2e_s^2)=\frac{1}{3\sqrt 6},\\
C^{(0)}=&\frac{1}{\sqrt 3} (e_u^2+e_d^2+e_s^2)=\frac{2}{3\sqrt 3}, 
\end{align} 	
the factors in front of the brackets stem from the definition of axial currents. The sum is over the flavors of light quarks $i=u,d,s$; $\lambda^a$ are the diagonal Gell-Mann $SU(3)$ matrices, $a=3,8$.
	
In particular, the anomaly sum rules for the isovector ($a=3$) and octet ($a=8$) currents when one of the photons is real ($k^2=0$) and another is real or virtual ($Q^2=-q^2 \geq 0$) read (in what follows we put $m_u=m_d=m_s=0$) \cite{Horejsi:1994aj, Klopot:2010ke}  
	\begin{equation}\label{asr38}
	\frac{1}{\pi}\int_{0}^{\infty}A^{(3,8)}_3(s,Q^2)ds = e^2\frac{C^{(3,8)}N_c}{2\pi^2},
	\end{equation}
	where the spectral density function is defined as $A^{(3,8)}_{3} \equiv \frac{1}{2}Im(F_3-F_6)$ and is determined from a decomposition of the vector-vector-axial (VVA) amplitude
	\begin{equation} \label{VVA}
	T_{\alpha \mu\nu}(k,q)=e^2\int
	d^4 x d^4 y e^{(ikx+iqy)} \langle 0|T\{ J_{\alpha 5}(0) J_\mu (x)
	J_\nu(y) \}|0\rangle
	\end{equation}
	as \cite{Rosenberg:1962pp} (see also \cite{Eletsky:1982py,Radyushkin:1996tb}) 
	\begin{align}
	\label{eq1} \nonumber T_{\alpha \mu \nu} (k,q)  & =  F_{1} \;
	\varepsilon_{\alpha \mu \nu \rho} k^{\rho} + F_{2} \;
	\varepsilon_{\alpha \mu \nu \rho} q^{\rho} + F_{3} \; k_{\nu} \varepsilon_{\alpha \mu \rho \sigma}k^{\rho} q^{\sigma}
	\\ 
	& + F_{4} \; q_{\nu} \varepsilon_{\alpha \mu \rho \sigma} k^{\rho}q^{\sigma} + F_{5} \; k_{\mu} \varepsilon_{\alpha \nu \rho \sigma} k^{\rho} q^{\sigma} + F_{6} \; q_{\mu} \varepsilon_{\alpha \nu \rho \sigma} k^{\rho} q^{\sigma},
	\end{align}
	where the coefficients $F_{j} = F_{j}(p^{2},q^{2})$, $j = 1, \dots ,6$ are the corresponding Lorentz invariant amplitudes constrained by current conservation and Bose symmetry. The electromagnetic currents are defined as $J_{\mu}=\sum_i {e_i\bar{\psi_i}\gamma_{\mu}\psi_i}, \ i=u,d,s$. 
	The r.h.s. of (\ref{asr38}) is exactly the Abelian (electromagnetic) anomaly constant stemmed from the matrix element $\langle 0 |F\tilde{F} |\gamma\gamma^* \rangle$. Note that \eqref{asr38} itself is a pure theoretical result obtained directly from dispersion representation of axial anomaly  \cite{Dolgov:1971ri, Horejsi:1985qu, Horejsi:1994aj}
	
	In the case of the strong anomaly term in the singlet current (\ref{an-0}), the analogous anomaly sum rule has an additional part stemmed from the matrix element 
	\begin{equation} \label{N}
	\langle 0 | \frac{\sqrt{3}\alpha_s}{4\pi} G\tilde{G}|\gamma\gamma \rangle = e^2C^{(0)}N_c N(p^2, k^2,q^2) \epsilon^{\mu\nu\rho\sigma}k_{\mu}q_{\nu}\epsilon_{\rho}^{(k)}\epsilon_{\sigma}^{(q)}.
	\end{equation}
\begin{figure}[H]
\caption{The matrix element $\langle 0 | \frac{\sqrt{3}\alpha_s}{4\pi} G\tilde{G}|\gamma\gamma \rangle$ schematic representation.}
\includegraphics[width=0.5\textwidth]{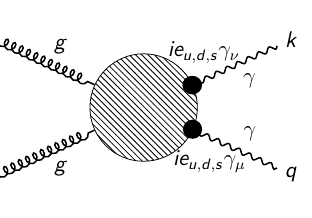}\label{figqed}
\end{figure}
This matrix element can be schematically illustrated by a diagram shown in the Fig.\ref{figqed}. The dashed circle denotes all possible perturbative and non-perturbative strong interactions in which finally one gets two photons coupled to the quark line with corresponding EM quark vertices $e_{u,d,s}$ for convenience shown by black circles. Thus one need to sum over all $u,d,s$ quark charges and so the matrix element \eqref{N} has the same charge factor coefficient $C^{(0)}$ as the matrix element $\langle 0 |F\tilde{F} |\gamma\gamma^* \rangle$. The factor $N_c$ is written explicitly for convenience. 

The corresponding non-Abelian contribution in the dispersive form requires a subtraction, so the singlet current ASR in the considered kinematics $N(p^2, k^2=0, q^2=-Q^2)\equiv N(p^2, Q^2)$ can be written  \cite{Khlebtsov:2018roy}, \cite{Khlebtsov:2020rte} as  
\begin{equation}\label{asr0}
\frac{1}{\pi}\int_{0}^{\infty}A^{(0)}_{3}(s,Q^{2})ds = \frac{e^2C^{(0)}N_c}{2\pi^2} + e^2C^{(0)}N_c\left(N(0,Q^2)-\frac{1}{\pi}\int_{0}^{\infty}Im R(s,Q^2)ds\right),
\end{equation}
where 
\begin{equation}\label{N_sub}
	R(p^2,Q^2)=\frac{1}{p^2}(N(p^2,Q^2)-N(0,Q^2)).
\end{equation}

Therefore, in addition to the Abelian anomaly contribution, the r.h.s. of Eq. (\ref{asr0}) has also a non-Abelian anomaly contribution (\ref{N}) given by the subtraction $N(0, Q^2)$ and the spectral parts. 

We saturate the l.h.s. of ASRs (\ref{asr38}) and (\ref{asr0}) with a full set of resonances and single out the lowest-lying contributing states in each channel in terms of the corresponding transition form factors and decay constants. The ``continuum" contribution absorbs the rest (higher resonances)  described by the same function $A_3(s,Q^2)$
\begin{equation}\label{qhd}
A^{(a)}_3(s,Q^2)=\pi  \Sigma f_P^{(a)}(s)\delta(s-m^{2}_P) F_{P\gamma}(s, Q^2)+A_3^{(a)}(s,Q^2) \theta(s-s_{a}),  \; a=3, 8, 0.
\end{equation}
Here the sum is over the hadron states $P$ whose decay constants $f_P^{(a)}$ and the form factors $F_{P\gamma}$ of the transitions $\gamma\gamma^* \to P$ are defined as
\begin{equation} \label{def_f}
\langle 0|J^{(a)}_{\alpha 5}(0) |P(p)\rangle=
i p_\alpha f^{(a)}_P(p^2), 
\end{equation}
\begin{equation} \label{def_tff}
\int d^{4}x e^{ikx} \langle P(p)|T\{J_\mu (x) J_\nu(0)
\}|0\rangle = e^2\epsilon_{\mu\nu\rho\sigma}k^\rho q^\sigma
F_{P\gamma}(p^2,Q^2)\;. 
\end{equation}
Thus the ASRs  (\ref{asr38}) and (\ref{asr0}) read (we omit factor $e^2$ for brevity),
\begin{equation}\label{asr38res}
\Sigma f_P^{(3,8)}(m_P^2)F_{P\gamma}(m_P^2,Q^2)+\frac{1}{\pi}\int_{s_{3,8}}^{\infty}A^{(3,8)}_3(s,Q^2)ds = \frac{C^{(3,8)}N_c}{2\pi^2},
\end{equation}
\begin{equation}\label{asr0res}
\Sigma f_P^{(0)}(m_P^2)F_{P\gamma}(m_P^2,Q^2)+\frac{1}{\pi}\int_{s_0}^{\infty}A_3^{(0)}ds = \frac{C^{(0)}N_c}{2\pi^2} + C^{(0)}N_c\left(N(0,Q^2)-\frac{1}{\pi}\int_{0}^{\infty}Im R(s,Q^2)ds\right).
\end{equation}  
	
	\begin{figure}[H]\label{fig:1}
		\ffigbox{
			\begin{subfloatrow}[3]
				\ffigbox[\FBwidth]{\caption{} \label{fig:2}}%
				{\includegraphics[width=0.33\textwidth]{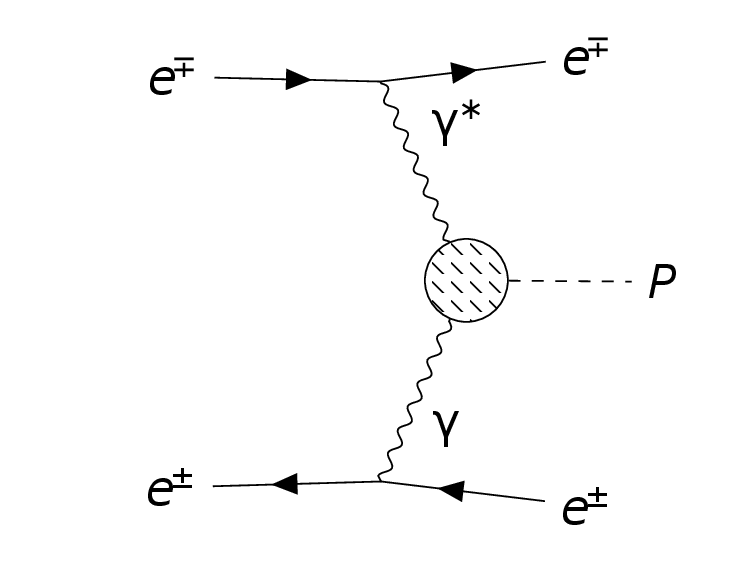}}
				\ffigbox[\FBwidth]{\caption{}\label{fig:3}}%
				{\includegraphics[width=0.33\textwidth]{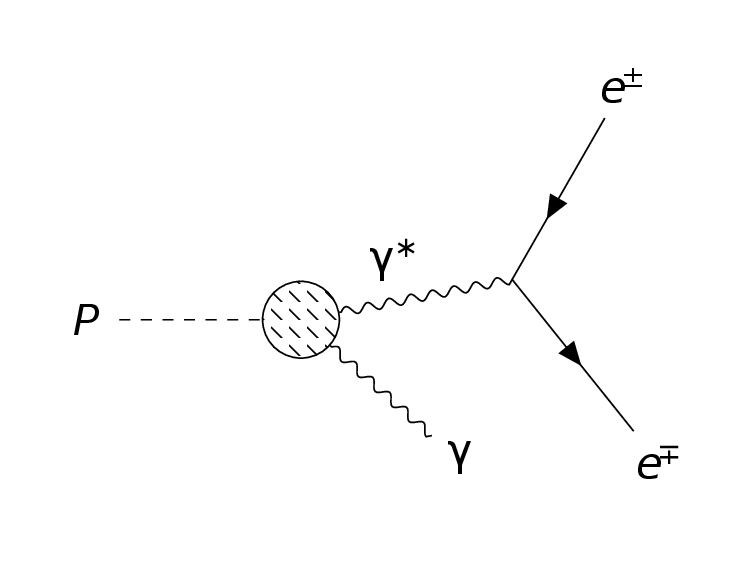}}  
				\ffigbox[\FBwidth]{\caption{} \label{fig:4}}%
				{\includegraphics[width=0.33\textwidth]{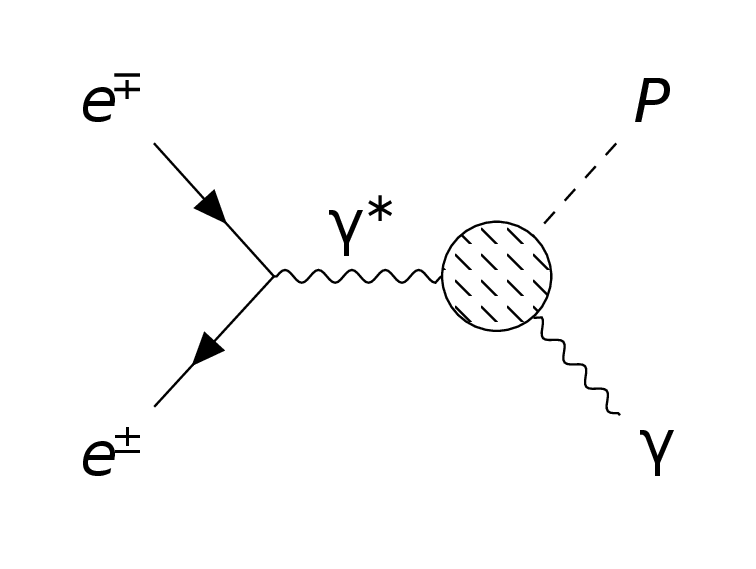}}
			\end{subfloatrow}
		}
		{
			\caption{The Feynman diagrams of the processes: (a)  $e^{+}e^{-}\rightarrow e^{+}e^{-}P$ scattering, (b) $P\rightarrow e^{+}e^{-}\gamma$ Dalitz decays, (c)  $e^{+}e^{-}\rightarrow P\gamma$ annihilation}
		}
	\end{figure}
	
	The lower integration limits $s_3,s_8,s_0$ in \eqref{asr38res}, \eqref{asr0res}, emerging as free parameters of the ASR approach,  strictly speaking, can be the functions of $Q^2$. Their values can be obtained from comparison with experiment or from comparison with other theoretical approaches.  Detailed discussion of their numerical values evaluation is presented in the paper \cite{Khlebtsov:2020rte}. It is natural to assume that the duality intervals of the isovector and octet currents cannot essentially differ: $s_8\simeq s_3$ within $20\%$ uncertainty of the $SU(3)$  symmetry braking.  For the purposes of numerical analysis, we take $s_8\simeq s_3 =0.6$ GeV$^2$ in this paper. The duality interval of the singlet current $s_0$ is different from $s_3$ and $s_8$, we take  $s_0 \simeq 1$ GeV$^2$.

\begin{figure}[H]
		\begin{floatrow}
			\ffigbox{\caption{$A_{QED}$.} \label{fig:5}}%
			{\includegraphics[width=0.5\textwidth]{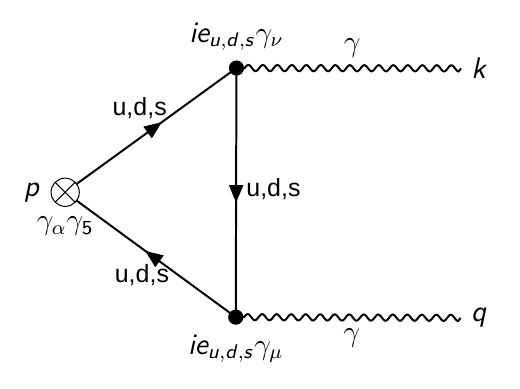}}
			\ffigbox{\caption{$A_{QCD}$.}\label{fig:6}}%
			{\includegraphics[width=0.5\textwidth]{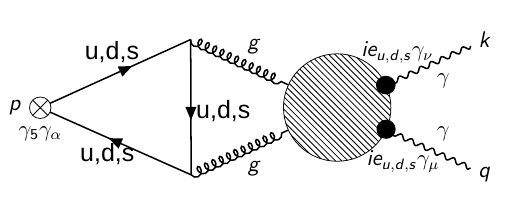}}  \end{floatrow}
	\end{figure}

	The one-loop approximation for the spectral densities of the isovector and octet currents $A_3^{(3,8)}(s,Q^2)$ (Fig. \ref{fig:5}) is given by \cite{Radyushkin:1996tb} 
	
	\begin{align} \label{a3}
	A_{3}^{(3,8)}(s, Q^2)=\frac{C^{(3,8)}N_c}{2\pi}\frac{Q^2}{(s+Q^2)^2},
	\end{align}
	so that the integration in the ASRs (\ref{asr38res}) leads to the following expressions for the hadron contributions,
	\begin{equation} \label{asr38aa}
	\Sigma f_P^{(3,8)}F_{P\gamma}(Q^2) = \frac{C^{(3,8)}N_c}{2\pi^2}\frac{s_{3,8}}{s_{3,8}+Q^2}.
	\end{equation}
	
	The case of the singlet current differs from the isovector and octet currents due to a new type of diagrams involving virtual gluons. 
	In order to single out electromagnetic contribution, we split the spectral density into two parts,
	\begin{equation}
	A_3^{(0)} = A_{QED}^{(0)}+A_{QCD}^{(0)}.
	\end{equation}
	The second part $A_{QCD}^{(0)}$ is the contribution of diagrams (Fig.~\ref{fig:6}) with virtual gluons coupled to two photons through all possible perturbative and non-perturbative strong interactions (see also \cite{Khlebtsov:2018roy}). The first part $A_{QED}^{(0)}$ represents the contribution of QED diagrams, whose lowest one-loop part (Fig. \ref{fig:5}) is given by a similar expression to Eq. (\ref{a3}) with an appropriate charge factor $C^{(0)}$. Making use of it, we can rewrite the ASR \eqref{asr0res} as
	\begin{equation}\label{qedplusqcd}
	\Sigma f_P^{(0)}F_{P\gamma}(Q^2) = \frac{N_cC^{(0)}}{2\pi^2}\frac{s_0}{s_0+Q^2} - C^{(0)}N_c(\frac{1}{\pi}\int_{s_0}^{\infty} A_{QCD}ds + N(0,Q^2) - \frac{1}{\pi}\int_{0}^{\infty} ImR(s,Q^{2})ds).
	\end{equation} 
	The first and the last three terms in Eq. (\ref{qedplusqcd}) represent the electromagnetic and the strong anomaly contributions to the ASR respectively. It is convenient to introduce a function that represents the ratio of contributions of strong and electromagnetic anomalies:
	\begin{equation}\label{gluon_anom_contrib1}
	B(Q^2, s_0) = \frac{2\pi^2}{N_cC^{(0)}}\frac{s_0+Q^2}{s_0}  \left[C^{(0)}N_c(N(0,Q^2) - \frac{1}{\pi}\int_{0}^{\infty} ImR(s,Q^{2})ds - \frac{1}{\pi}\int_{s_0}^{\infty} A_{QCD}(s,Q^2)ds) \right].
	\end{equation} 
	
	As the integral of $A_{QCD}$ is suppressed as $\alpha^2_s$ at $s_0 \geq 1.0$ GeV$^2$, the function $B(Q^2,s_0)$ is predominantly determined by the first two terms. It reflects the properties of the non-perturbative matrix element  $\langle 0 |G\tilde{G}|\gamma\gamma^{(*)} \rangle$. 
Therefore, the study of the function $B(Q^2,s_0)$ gives us access to the non-Abelian anomaly contribution to the processes $P\to \gamma^*\gamma$ at various photon virtualities.

Let us note, that the charge factor coefficients of the matrix elements of strong and electromagnetic anomalies appear to be the same. So the function $B(Q^2,s_0)$, as it is a ratio of them, does not depend on vector vertices  couplings.
	
	So, rewriting the ASR for the singlet current \eqref{qedplusqcd} in terms of the function $B(Q^2,s_0)$ \eqref{gluon_anom_contrib1}, one gets
	
	\begin{equation}\label{mix0}
	\Sigma f_P^{(0)}F_{P\gamma}(Q^2) = \frac{N_cC^{(0)}}{2\pi^2}\frac{s_0}{s_0+Q^2} \left[1+B(Q^2, s_0) \right].
	\end{equation}
	
	Taking into account the lowest contributions, given by the $\pi^0, \eta$ and $\eta'$ mesons, the ASRs for the isovector, octet (\ref{asr38aa}) and singlet (\ref{mix0}) currents comprise a system of equations,
	\begin{equation}\label{system}
	\left(
	\begin{matrix}
	f_{\pi^0}^{(3)} & f_{\eta}^{(3)} & f_{\eta'}^{(3)} \\
	f_{\pi^0}^{(8)} & f_{\eta}^{(8)} & f_{\eta'}^{(8)} \\
	f_{\pi^0}^{(0)} & f_{\eta}^{(0)} & f_{\eta'}^{(0)} 
	\end{matrix}
	\right)
	\left(
	\begin{matrix}
	F_{\pi^0}(Q^2) \\
	F_{\eta}(Q^2)  \\
	F_{\eta'}(Q^2)
	\end{matrix}
	\right)=
	\left(
	\begin{matrix}
	\frac{N_cC^{(3)}}{2\pi^2}\frac{s_3}{s_3+Q^2} \\
	\frac{N_cC^{(8)}}{2\pi^2}\frac{s_8}{s_8+Q^2}  \\
	\frac{N_cC^{(0)}}{2\pi^2}\frac{s_0(1+B(Q^2,s_0))}{s_0+Q^2}
	\end{matrix}
	\right),
	\end{equation}
	whose solution leads \cite{Khlebtsov:2020rte} to the expressions for the form factors,
	
	\begin{equation}\label{solution_generic}
	F_{P\gamma}(Q^2) = \alpha_{P} \frac{s_3}{s_3+Q^2} + \beta_{P} \frac{s_8}{s_8+Q^2} + \gamma_{P} \frac{s_0}{s_0+Q^2}[1+B(Q^2,s_0)],
	\end{equation}
	where $P = \pi^0, \eta, \eta'$.  The coefficients $\alpha_P$, $\beta_P$, $\gamma_P$ are expressed in terms of the decay constants $f_P^{(i)}$,  $\Delta$ is the determinant of the decay constant matrix in (\ref{system}):
	\begin{align}
	\small &\alpha_{\pi^0}= \frac{N_cC^{(3)}}{2\pi^2\Delta}(f^{(8)}_{\eta}f^{(0)}_{\eta'} - f^{(0)}_{\eta}f^{(8)}_{\eta'}),&\beta_{\pi^0}= \frac{N_cC^{(8)}}{2\pi^2\Delta}(f^{(0)}_{\eta}f^{(3)}_{\eta'} - f^{(3)}_{\eta}f^{(0)}_{\eta'}),&\gamma_{\pi^0}= \frac{N_cC^{(0)}}{2\pi^2\Delta}(f^{(3)}_{\eta}f^{(8)}_{\eta'} - f^{(8)}_{\eta}f^{(3)}_{\eta'}), \label{coef_pi}\\ \label{coef_eta}
	&\alpha_{\eta}= \frac{N_cC^{(3)}}{2\pi^2\Delta}(f^{(0)}_{\pi^0}f^{(8)}_{\eta'} - f^{(8)}_{\pi^0}f^{(0)}_{\eta'}),&\beta_{\eta}= \frac{N_cC^{(8)}}{2\pi^2\Delta}(f^{(3)}_{\pi^0}f^{(0)}_{\eta'} - f^{(0)}_{\pi^0}f^{(3)}_{\eta'})\;,&\gamma_{\eta}= \frac{N_cC^{(0)}}{2\pi^2\Delta}(f^{(8)}_{\pi^0}f^{(3)}_{\eta'} - f^{(3)}_{\pi^0}f^{(8)}_{\eta'}),\\ \label{coef_etaprime}
	&\alpha_{\eta'}= \frac{N_cC^{(3)}}{2\pi^2\Delta}(f^{(8)}_{\pi^0}f^{(0)}_{\eta} - f^{(0)}_{\pi^0}f^{(8)}_{\eta}),&\beta_{\eta'}= \frac{N_cC^{(8)}}{2\pi^2\Delta}(f^{(0)}_{\pi^0}f^{(3)}_{\eta} - f^{(3)}_{\pi^0}f^{(0)}_{\eta}),&\gamma_{\eta'}= \frac{N_cC^{(0)}}{2\pi^2\Delta}(f^{(3)}_{\pi^0}f^{(8)}_{\eta} - f^{(8)}_{\pi^0}f^{(3)}_{\eta}).
	\end{align}
		So, the expressions for the TFFs \eqref{solution_generic} are the consequence of the dispersive approach to axial anomaly comibined with quark-hadron duality. In this way, it provides  theoretical grounds for the Brodsky-Lepage interpolation formula for pion TFF \cite{Brodsky:1981rp} and to some of its generalizations to the $\eta$ and $\eta'$ TFFs \cite{Feldmann:1998yc}. 

For the purposes of numerical study, we use several decay constant sets, obtained in different analyses of the mixing parameters. Their numerical values are listed in paper \cite{Khlebtsov:2020rte}. The corresponding values of $\alpha_{P},\beta_{P},\gamma_{P}$  for different decay constants sets are shown in Table \ref{table1}. 
	
\begin{table}[H]
\centering
		\caption{The coefficients $\alpha_{P},\beta_{P},\gamma_{P}$ \eqref{coef_pi}, \eqref{coef_eta}, \eqref{coef_etaprime} in GeV$^{-1}$ from different analyses of the mixing parameters. }
		\label{table1}
\begin{tabular}{l|c|c|c|c|c|c|c|c|c}
\hline
\multirow{2}{*}{Mix. sch.} & \multicolumn{3}{c|}{$\pi^0$} & \multicolumn{3}{c|}{$\eta$} & \multicolumn{3}{c}{$\eta'$} \\ \cline{2-10} 
			& $\alpha_{\pi^0}$ & $\beta_{\pi^0}$ & $\gamma_{\pi^0}$ & $\alpha_{\eta}$ & $\beta_{\eta}$ & $\gamma_{\eta}$ & $\alpha_{\eta'}$ & $\beta_{\eta'}$  & $\gamma_{\eta'}$    \\ \hline
  FKS98\cite{Feldmann:1998vh} & 0.274 & -0.0005 & 0.013 & -0.0015 & 0.127 & 0.144 & -0.0078 & -0.021 & 0.365 \\
  EF05\cite{Escribano:2005qq} & 0.274 & $4\times 10^{-6}$ & 0.012 & -0.0017 & 0.112 & 0.145 & -0.0071 & -0.0047 & 0.341 \\ 
  KOT12\cite{Klopot:2012hd} & 0.274 & -0.0005 & 0.014 & -0.0016 & 0.135 & 0.154 & -0.0087 & -0.021 & 0.406 \\ 
  EGMS16\cite{Escribano:2015yup} & 0.274 & -0.0004 & 0.013 & -0.0016 & 0.128 & 0.147 & -0.008 & -0.016 & 0.377 \\ 
\end{tabular}
\end{table}	

For the case of two real photons it was established that the function $B(Q^2,s_0)$ value \eqref{gluon_anom_contrib1} is close to zero, the results for all considered mixing schemes listed in the Table \ref{table2}.
\begin{table}[H]
\caption{The values of $B(0)$ in various mixing schemes tacking into account mixing between $\pi^0$ and $\eta-\eta'$.}\label{table2}
\begin{tabular}{c|c|c|c|c}
   \hline
    Mix.scheme & FKS98\cite{Feldmann:1998vh} & EF05\cite{Escribano:2005qq} & KOT12\cite{Klopot:2012hd} & EGMS16\cite{Escribano:2015yup} \\ \hline
    B(0) & 0.022 & 0.045 & -0.080 & -0.024  \\ 
\end{tabular}
\end{table}

\begin{figure}[H]
	\caption{\label{fig:11}
		Contour plot of $\chi^2_{\eta+\eta'}(B,s_0)/dof$ for $\eta, \eta'$ TFFs and the data from Refs. \cite{Acciarri:1998,Behrend:1990sr,Gronberg:1997fj,BABAR:2011ad} using the EGMS16\cite{Escribano:2015yup} mixing scheme. Dashed black contour and black filled circle denote the $99\%$ C.L. line and the minimum respectively for the combined $\eta +\eta'$ data;  the black filled square denotes $\chi^2$ minimum for the $\eta'$ data.
	}
	\includegraphics[width=0.5\textwidth]{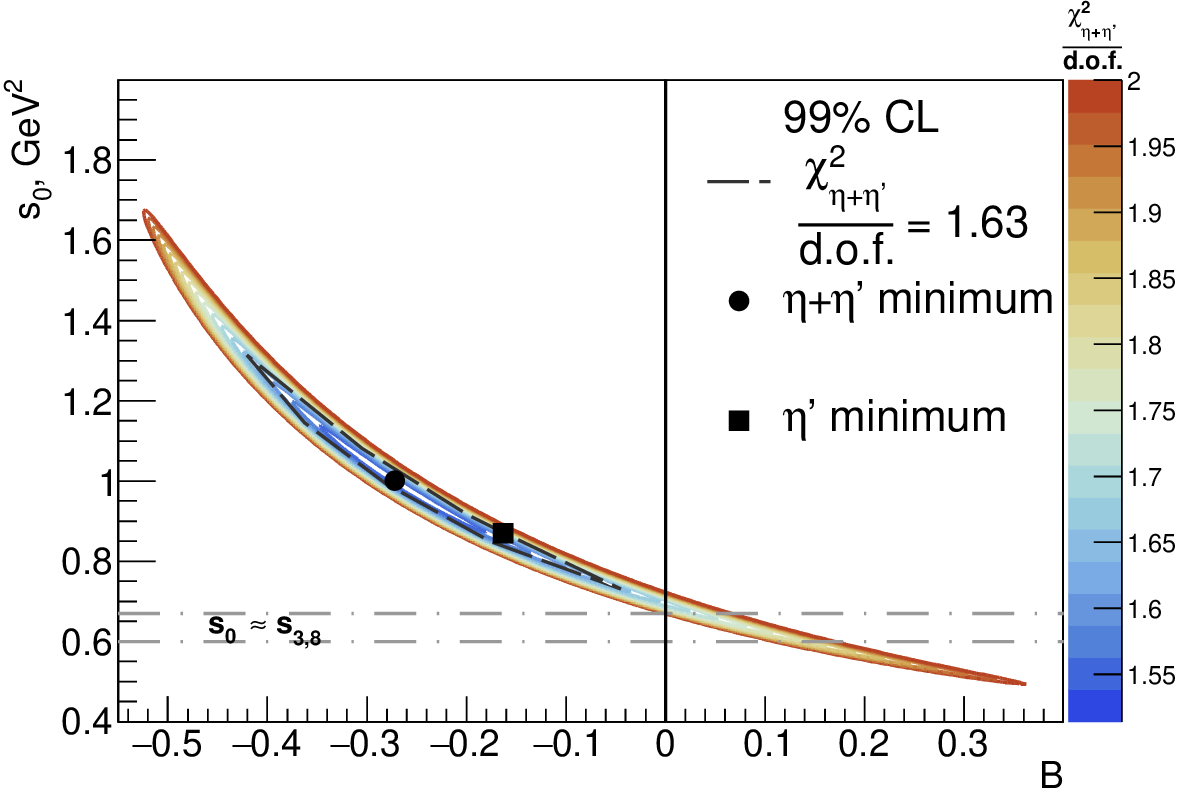}
\end{figure}

In the space-like domain  the values of $B(Q^2,s_0)$ at $Q^2>0.6$ GeV$^2$ (the available data at $Q^2<0.6$ are scarce and have large uncertainties)  considerably differ from $B(0)\sim 0$ at $Q^2=0$ for all considered mixing schemes, as seen in the Table \ref{table_s_and_B}. Also note that the values  $B_{as}-B(0)$ appear to be close in different mixing schemes.  
\begin{table}[H]
\caption{The values of $s_0$ and $B$ for different mixing schemes from two-parameter fit (with the statistical errors).
}\label{table_s_and_B}
\begin{tabular}{c|c|c|c|c}
   \hline
    Mix. sch. & $s_0$ & $B_{as}$ & $B_{as} - B(0)$ & $\frac{\chi^2_{\eta+\eta'}}{dof=81-2}$  \\ \hline
    FKS98\cite{Feldmann:1998vh} & $1.00(10)$& $-0.242(62)$ & -0.264 & 1.66 \\ 
    EF05\cite{Escribano:2005qq} & $0.99(10)$ & $-0.209(66)$ & -0.254 & 1.20  \\ 
    KOT12\cite{Klopot:2012hd} & $1.01(10)$ & $-0.320(56)$ & -0.240  & 1.74 \\ 
    EGMS16\cite{Escribano:2015yup} & $1.00(10)$ & $-0.272(60)$ & -0.248  & 1.51 \\ 
\end{tabular}
\end{table}
The contour plot of $\chi^2_{\eta+\eta'}/dof$ for the mixing parameters EGMS16\cite{Escribano:2015yup} is shown in Fig. \ref{fig:11}.  The results for other mixing schemes are similar.
The  black filled circle and the black filled square indicate the minima of $\chi^2$ from the combined $\eta+\eta'$ and separate $\eta'$ data fits respectively; they show rather good agreement with each other.
This plot demonstrates a correlation between $B$ and $s_0$. In particular, the best-fit values are $s_0\simeq 1$ GeV$^2$, $B\simeq-0.25$.
At the same time it shows a possibility when $B\sim 0$ and $s_0 \simeq s_3\simeq 0.6$~GeV$^2$, which means  a negligible role of the strong anomaly at larger $Q^2$, not only at $Q^2=0$. We will call this scenario ($s_0\simeq 0.6$ GeV$^2$ and $B(Q^2)\simeq 0$) a \textit{hidden strong anomaly} case, as opposite to the \textit{open strong anomaly} case ($s_0\simeq 1$~GeV$^2$ and $B(Q^2)\simeq -0.25$). Further study in the time-like region will show that these regimes correspond to different $q^2$ regions.

{The hidden anomaly scenario has the following physical interpretation. The mass of the dominant contributor to the singlet ASR, the $\eta'$ meson, originates from the strong anomaly contribution the famous ($U(1)_A$ problem) \cite{tHooft:1986ooh,Diakonov:1981nv}. Neglecting the strong anomaly brings the value of $s_0$  close to the values of the octet and isovector intervals of duality $s_8$ and $s_3$, dominated by the $\eta$ and $\pi^0$ mesons. One can say that strong anomaly contribution $B$ is "absorbed" by duality interval resulting in its decrease down to the "Abelian" value.

We may conclude, that at $Q^2\gtrsim 0.6$ GeV$^2$ the strong anomaly comprises $\sim 25\%$ of the electromagnetic anomaly contribution and is almost independent of the photon virtuality, while at smaller $Q^2$ it rapidly vanishes. A simple approximation of the function $B(Q^2)$ by a  step-function:
\begin{equation}\label{B_simple}
B(Q^2) = \begin{cases} \approx0, & \mbox{if } 0<Q^2<0.6 \ GeV^2, \\ B_{as}, & \mbox{if } Q^2>0.6 \ GeV^2,   \end{cases}
\end{equation}
The $\eta$ and $\eta'$ TFFs for different mixing parameters with step-function approximation of $B(Q^2)$ \eqref{B_simple}  is compared with the data in Figs.~\ref{fig:9} and \ref{fig:10}. The green shaded area shows the $20\%$ uncertainty of $s_8$ (for the EGMS16 \cite{Escribano:2015yup} mixing scheme).
\begin{figure}[H]
 \begin{floatrow}
   \ffigbox{\caption{The $\eta\rightarrow \gamma\gamma^*$ TFF \eqref{solution_generic} in the space-like region for different mixing schemes with the approximation of $B(Q^2)$ as a step function.  The insert shows a low $Q^2$ region. The green shaded area shows the $20\%$ uncertainty of $s_8$ (for the EGMS16 \cite{Escribano:2015yup} mixing scheme). The experimental data \cite{Behrend:1990sr, Gronberg:1997fj, BABAR:2011ad} are used.} \label{fig:9} .} %
           {\includegraphics[width=0.5\textwidth]{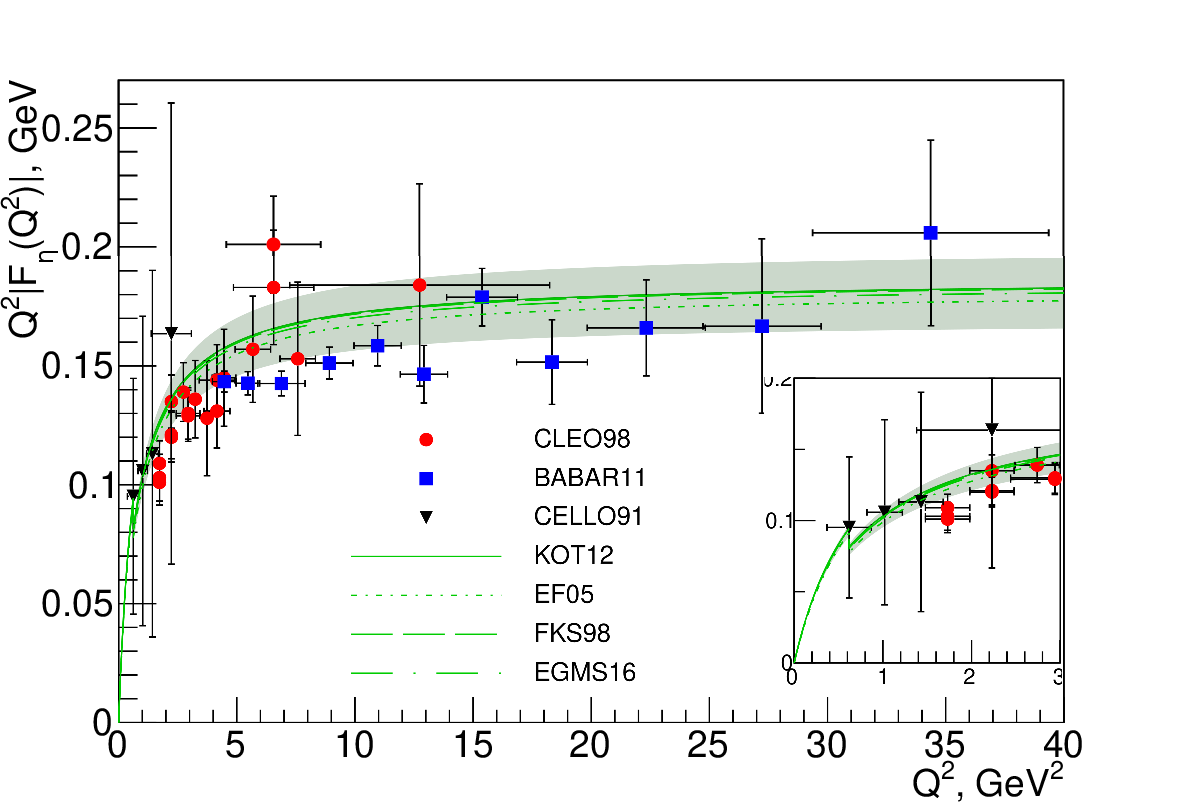}}
   \ffigbox{\caption{The $\eta'\rightarrow \gamma\gamma^*$ TFF \eqref{solution_generic} in the space-like region for different mixing schemes with the approximation of $B(Q^2)$ as a step function.  The insert shows a low $Q^2$ region. The experimental data  \cite{Acciarri:1998, Behrend:1990sr, Gronberg:1997fj, BABAR:2011ad} are used.}\label{fig:10}}%
           {\includegraphics[width=0.5\textwidth]{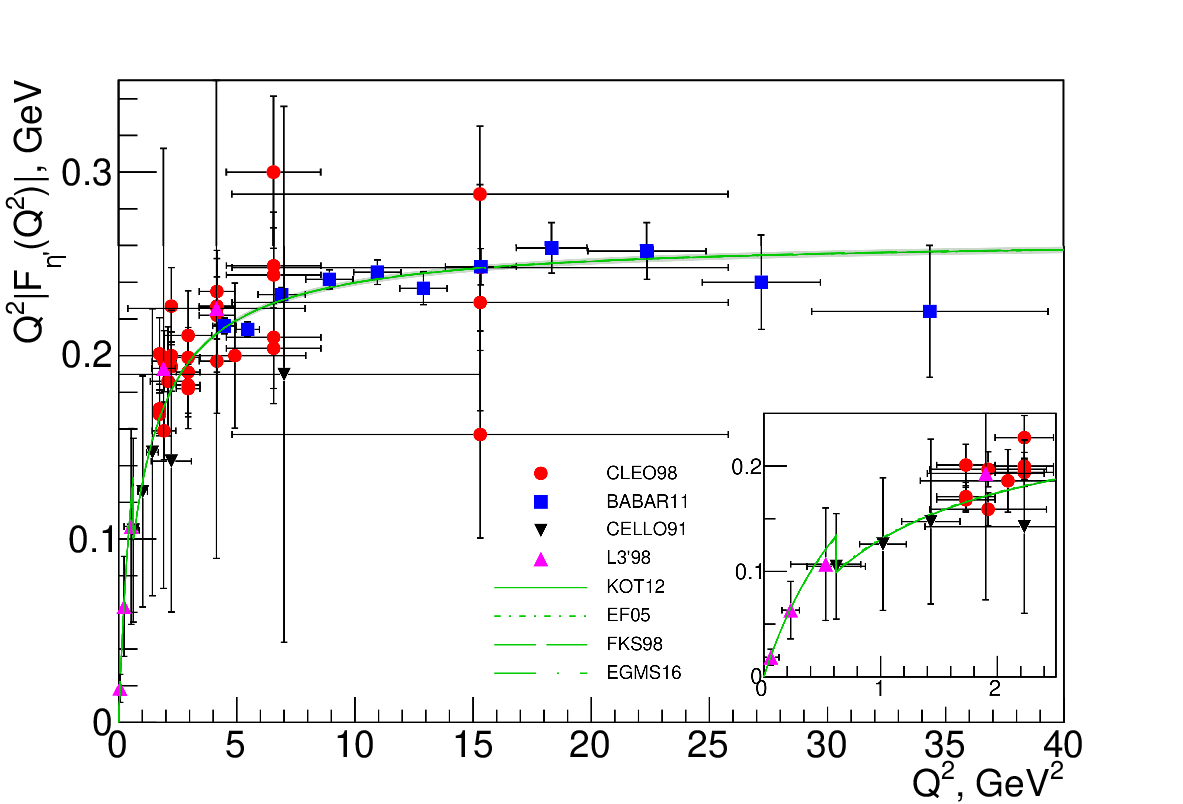}}         
  \end{floatrow}
\end{figure}

In the paper \cite{Khlebtsov:2020rte} it was shown that  ASR \eqref{asr38aa} and \eqref{mix0} can be analytically continued to the time-like region, and one can obtain equations for TFFs in this domain too
\begin{equation}\label{solution_generic_time}
|F_{P}(q^2)| = \left | \alpha_{P} \frac{s_3}{s_3-q^2} + \beta_{P} \frac{s_8}{s_8-q^2} + \gamma_{P} \frac{s_0}{s_0-q^2}\left[1+B(q^2,s_0)\right] \right |,
\end{equation}
where $P = \pi^0, \eta, \eta'$.

Let us stress out that the TFFs equations \eqref{solution_generic} and \eqref{solution_generic_time} are obtained by the axial anomaly dispersive representation and by the application of global quark-hadron duality to ASRs of the corresponding axial currents without any model considerations. In paper \cite{Khlebtsov:2020rte} it was pointed out that \eqref{solution_generic} and \eqref{solution_generic_time}  appear to be similar to the expresions obtained in the framework  of Vector Meson Dominance model \cite{Sakurai:1960ju,OConnell:1995nse}. 

We observe  correlation between the quantities inherent to the axial channel (and pseudoscalar hadron resonances) and vector hadron resonances: $s_3, s_8, s_0\longleftrightarrow m_{\rho}^2, m_{\omega}^2, m_{\phi}^2$. The decay constants of pseudoscalar mesons (which determine $\alpha_P,\beta_P,\gamma_P$ coefficients) correspond to the vector meson coupling constants (residues) of the VMD model. This means, in particular, that the large $\eta$-$\eta'$ mixing in the pseudoscalar sector is correlated with the residues of the vector mesons.

These observations confirm that the axial anomaly in its dispersive form (i.e. respective ASRs) reveals the duality between axial and vector channels, seen already \cite{Horejsi:1994aj} in perturbative calculations.  This duality may be also related \cite{Pasechnik:2005ae} to the theorems \cite{Vainshtein:2002nv} for longitudinal and transverse parts of two-point V-A correlators in external electromagnetic fields. There are also relations between resonances in different channels in the holographic approach \cite{Son:2010vc}.

In the Dalitz decay domain it was observed that hidden anomaly scenario is realised. The value of $s_0$ should be close to $s_{3,8}\approx0.6$ GeV$^2$. Also it was found that for the $s_8$ it is preferable to use value $\approx0.48$ GeV$^2$ to have better description of data. The results for $\eta$ and $\eta'$ TFFs \eqref{solution_generic_time} are shown in Fig. \ref{fig:14} and \ref{fig:15}.
\begin{figure}[H]
	\begin{floatrow}
		\ffigbox{\caption{The $\eta$ TFFs \eqref{solution_generic_time}. The red shaded area shows the $20\%$ uncertainty of $s_8$ (for the EGMS16 \cite{Escribano:2015yup} mixing scheme). The experimental data are by A2  \cite{Adlarson:2016hpp} and NA60 \cite{Arnaldi:2016pzu} collaborations.} \label{fig:14}}%
		{\includegraphics[width=0.5\textwidth]{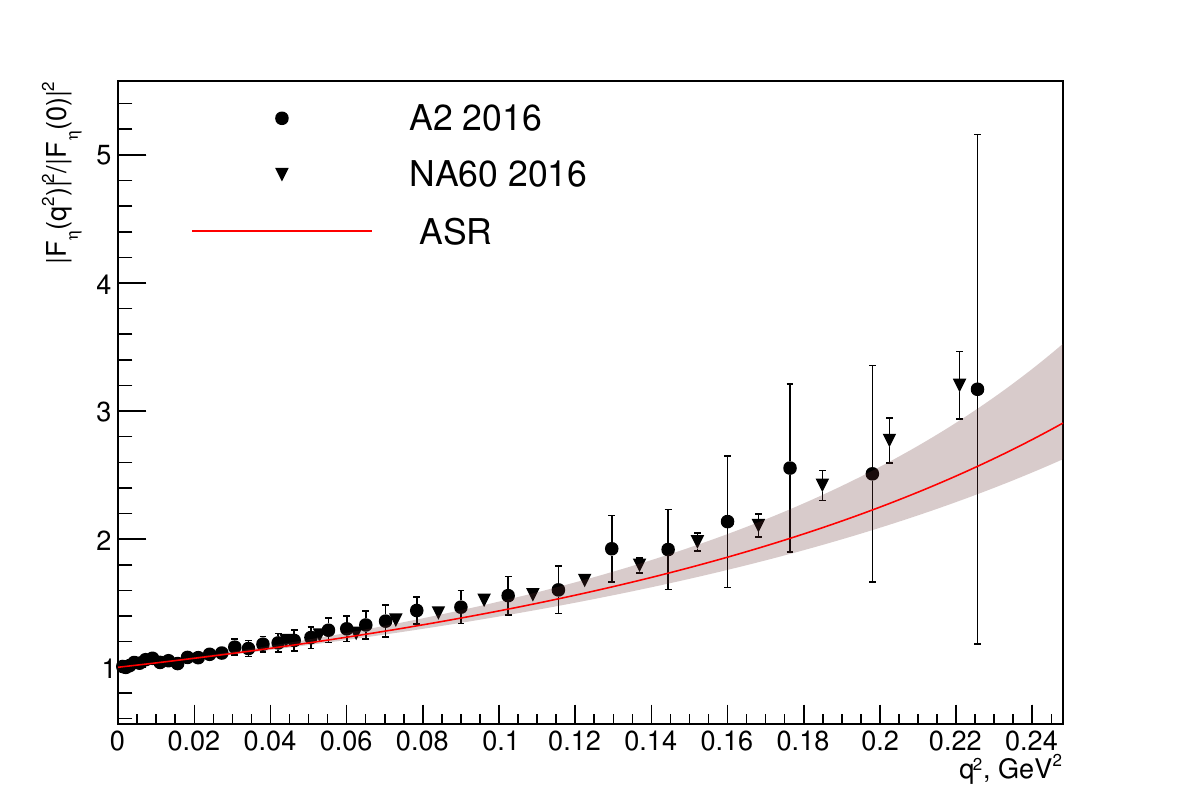}}
		\ffigbox{\caption{The $\eta'$ TFFs \eqref{solution_generic_time}. The experimental data are by BESIII collaboration \cite{Ablikim:2015wnx}. }\label{fig:15}}%
		{\includegraphics[width=0.5\textwidth]{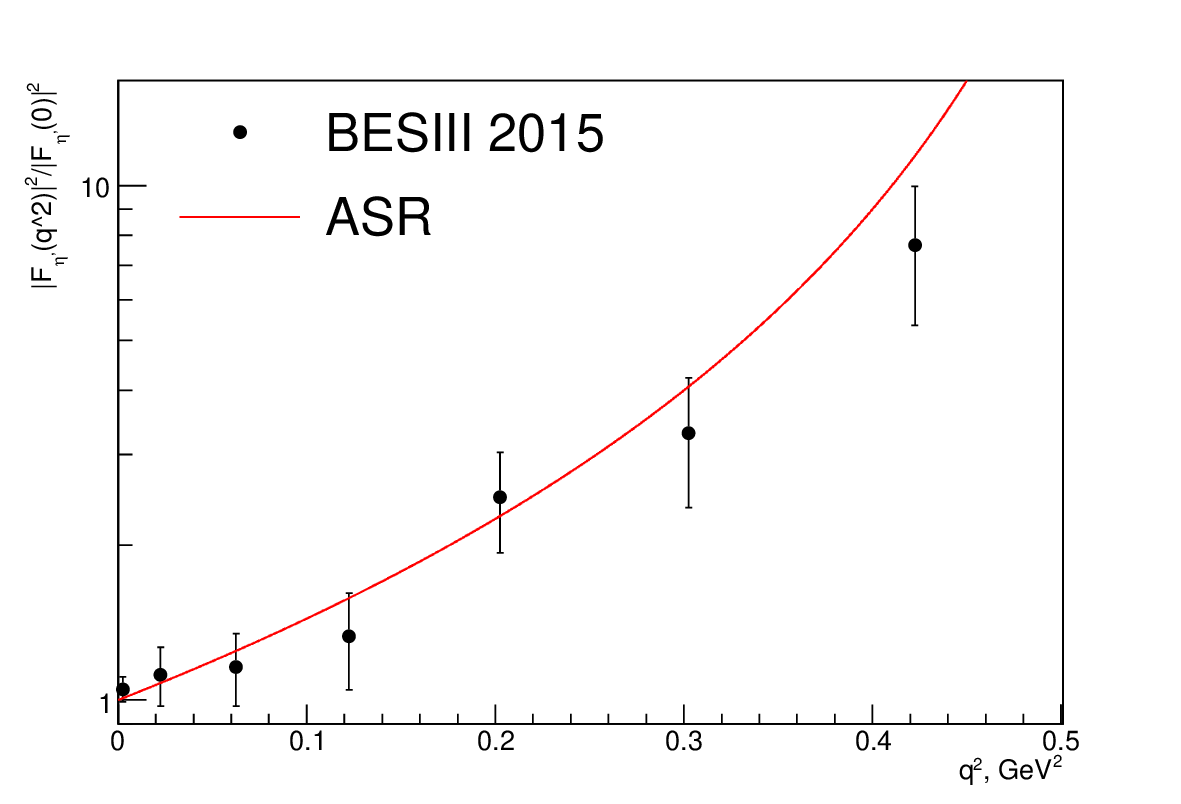}}         
	\end{floatrow}
\end{figure}

In the annihilation domain ($q^2>m^2_{P}$), we established that the pion TFF experimental data in the time-like region confirms the existence of $\pi^0-\eta-\eta'$ mixing, which means that the the pion TFF equation must have three terms.

We found that the function $B(q^2)$ should have a sharp minimum at $q^2\simeq 1$ GeV$^2$ with $B(q^2\simeq 1)<-1$. As a modification of $B(q^2)$, it was proposed a Gaussian function
\begin{equation}\label{B_modified}
B(q^2) = \begin{cases} 0, & \mbox{if } 0<q^2<0.6 \ GeV^2, \\ be^{-\frac{(q^2 -\mu)^2}{2c^2}} + B_{as}, & \mbox{if } q^2>0.6 \ GeV^2,   \end{cases}
\end{equation}
where $B_{as} = -0.262$ is the asymptotic value, and $b, \mu, c$ are the parameters.  The  $\pi^0$ and $\eta$ TFFs with $b=-1.1, c = 0.3$ GeV$^2$, $\mu = 1.0$ GeV$^2$ for the EGMS16 \cite{Escribano:2015yup} mixing scheme are shown in Fig. \ref{fig:14} and \ref{fig:15} as a solid red curve, respectively.
\begin{figure}[H]
	\begin{floatrow}
		\ffigbox{\caption{$\pi^0$ TFF \eqref{solution_generic_time} for the EGMS16 \cite{Escribano:2015yup} mixing scheme compared with the experimental data \cite{Adlarson:2016ykr,Achasov:2016bfr,Akhmetshin:2004gw,Achasov:2018ujw}. The solid red curve - TFF with a Gaussian $B(q^2)$ \eqref{B_modified}, the dot-dashed blue curve -- mixing of $\pi^0$ and $\eta-\eta'$ is neglected. The insert shows the Dalitz decay domain.} \label{fig:14}}%
		{\includegraphics[width=0.5\textwidth]{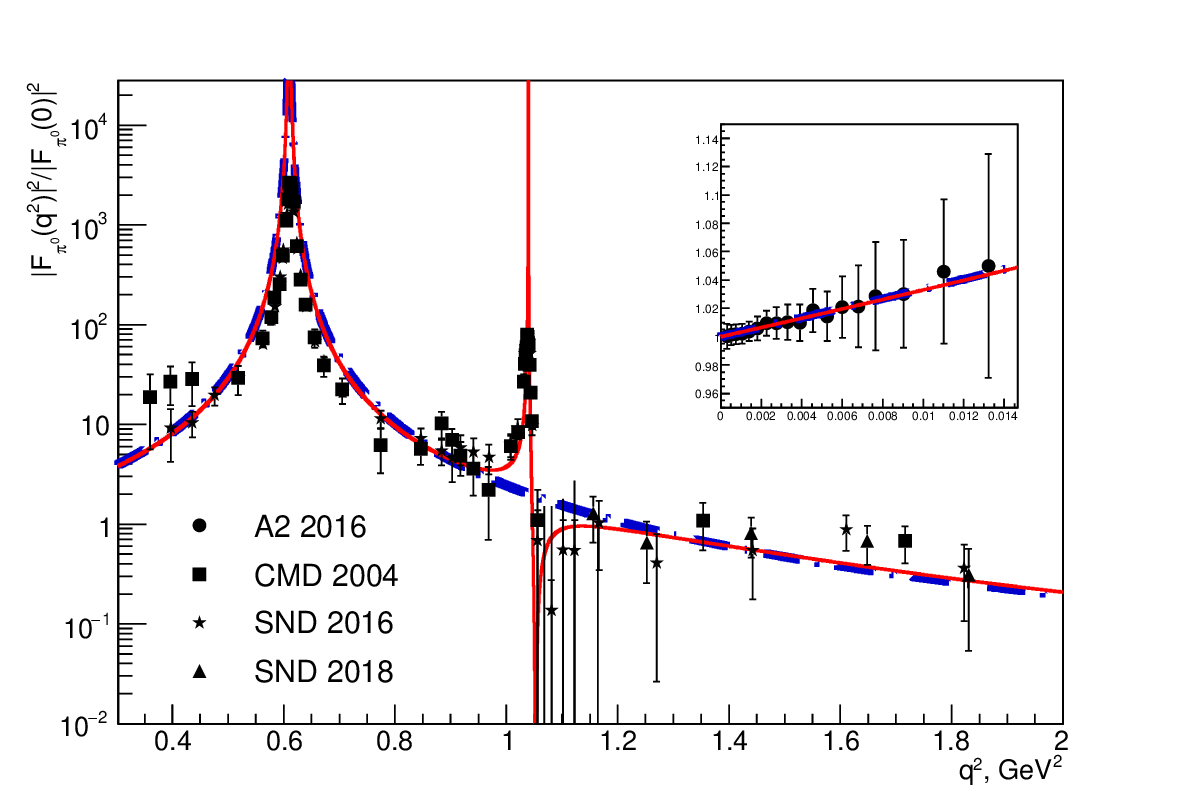}}
		\ffigbox{\caption{The $\eta$ TFF \eqref{solution_generic_time} for the EGMS16 \cite{Escribano:2015yup} mixing scheme compared with the experimental data \cite{Achasov:2007kw,Achasov:2013eli,Akhmetshin:2004gw}: the solid red curve -- $\eta$ TFF with Gaussian modification of $B(q^2)$ \eqref{B_modified}.}\label{fig:15}}%
		{\includegraphics[width=0.5\textwidth]{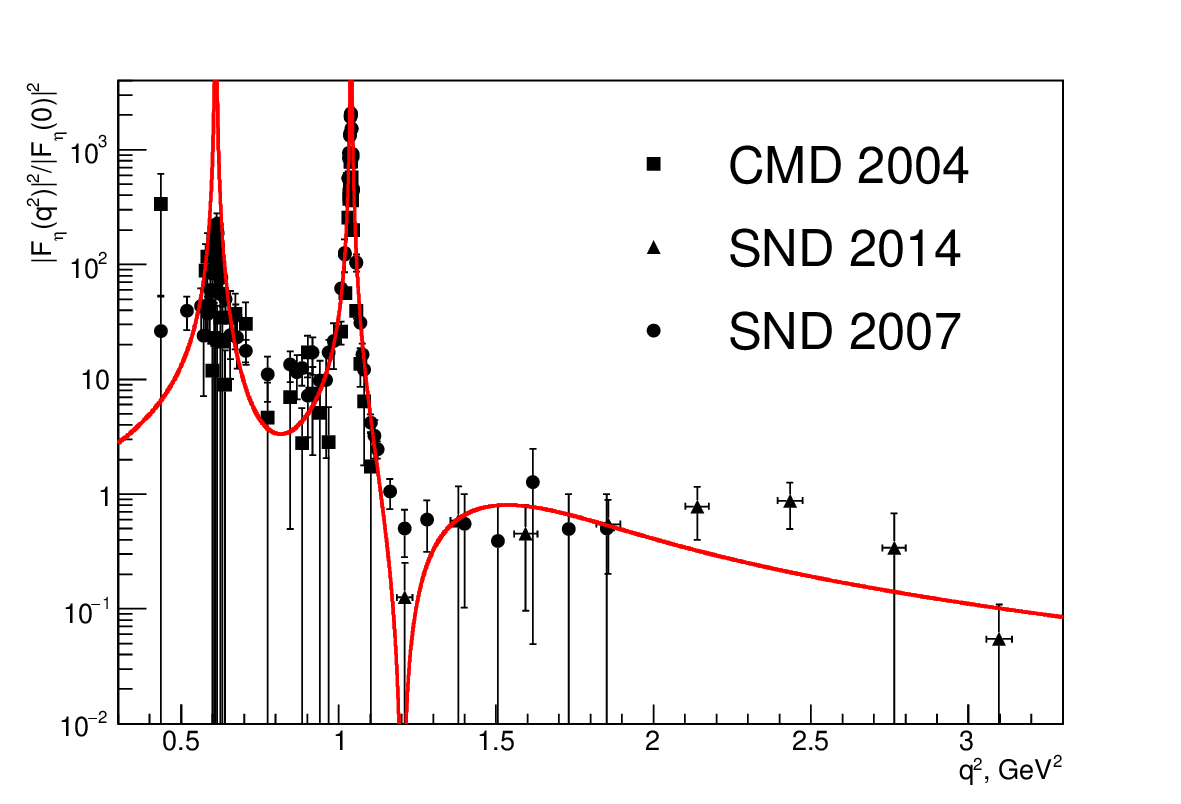}}         
	\end{floatrow}
\end{figure}

Thus the proposed $B(q^2)$ with Gaussian minimum at $q^2\simeq 1$ GeV$^2$ \eqref{B_modified} gives a consistent description of the $\pi^0, \eta$ TFFs in a wide range of $q^2$, providing correct reproduction of the data at small and large $q^2$ as well as the interference pattern near the poles.

Also note that the $\eta$ and $\eta'$ TFFs \eqref{solution_generic_time} are in a good agreement with the high-$q^2$ time-like region data measured at the point $q^2=112$ GeV$^2$ by BaBar \cite{Aubert:2006cy}.
\begin{table}[H]
\caption{$\eta$ and $\eta'$ \eqref{solution_generic_time} TFFs at $q^2 = 112$ GeV$^2$.}\label{table6}
\begin{tabular}{c|c|c}
	Mix. sch. &$q^2|F_{\eta}(q^2)|$, GeV &  $q^2|F_{\eta'}(q^2)|$, GeV  \\ \hline
    FKS98 \cite{Feldmann:1998vh} & 0.183 & 0.255 \\ 
	EF05 \cite{Escribano:2005qq} & 0.179 & 0.258 \\ 
	KOT12 \cite{Klopot:2012hd} & 0.184 & 0.256 \\ 
	EGMS16 \cite{Escribano:2015yup} & 0.183 & 0.262 \\
	\hline
	Experiment \cite{Aubert:2006cy} & 0.229(31) & 0.251(20) \\ 
\end{tabular}
\end{table}

Therefore, from the analysis of the $\pi^0$, $\eta$ and $\eta'$ meson TFFs at various photon virtualities, we can make the following conclusions for the strong to electromagnetic anomaly ratio $B(q^2)$. At low-$|q^2|$, the hidden (strong) anomaly case ($B\simeq0$ and $s_{0}\approx s_{8,3}$) takes place. At larger $|q^2|\gtrsim0.6$~GeV$^2$, the strong anomaly contribution to the TFFs rapidly becomes significant, reaching $\simeq 25\%$ of the electromagnetic one ($B\simeq-0.25$, $s_0\simeq1$~GeV$^2$). The function $B(q^2)$ has an extremum at the time-like $q^2 \simeq 1$ GeV$^2$.  Qualitatively, the function $B(q^2)$ can be described by a curve shown in Fig. \ref{fig:17}.

\begin{figure}[H]
	\caption{The strong/electromagnetic anomaly ratio function -- $B(q^2)$.} \label{fig:17}%
	{\includegraphics[width=0.5\textwidth]{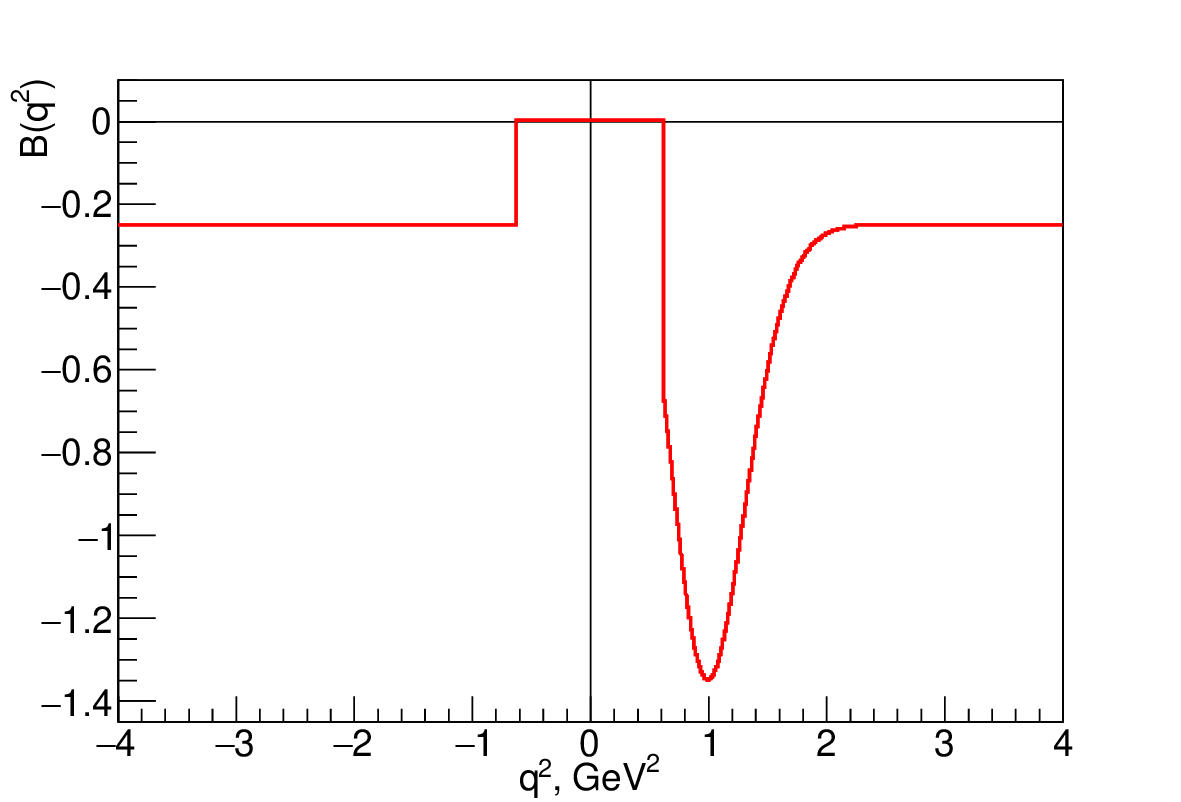}}
\end{figure}

The function $B(q^2)$ represented in the Fig. \ref{fig:17} demonstrates key features of the non-perturbative matrix element  $\langle 0 |G\tilde{G}|\gamma\gamma^{(*)} \rangle$, namely close to 0 value at $q^2=0$, the assymptoics in both space- and time-like domains and extremum at $q^2\approx1$ GeV$^2$. 

As pointed earlier in text the lower integration limits $s_3,s_8,s_0$ in \eqref{asr38res}, \eqref{asr0res},(appearing also in the equations \eqref{solution_generic} and \eqref{solution_generic_time}) emerging as free parameters of the ASR approach, can be functions of $q^2$. From the analysis of the $\pi^0$, $\eta$ and $\eta'$ meson TFFs at various kinematic domains we observed that they are indeed functions of $q^2$.

The $s_0(q^2)$ at low-$q^2$ should have value $\approx0.6$ GeV$^2$, while starting from $|q^2|\gtrsim0.6$~GeV$^2$ it should be $1$ GeV$^2$. So a simple approximation as step-function can be introduced, similar to \eqref{B_simple}, namely
\begin{equation}\label{s_0_simple}
s_0(q^2) = \begin{cases} \approx0.6 \ GeV^2, & \mbox{if } 0<|q^2|<0.6 \ GeV^2, \\ \approx1 \ GeV^2, & \mbox{if } |q^2|>0.6 \ GeV^2,   \end{cases}
\end{equation}

For the $s_8(q^2)$ it is also needed similar step-function approximation. As one can see in the Fig.\ref{fig:14} in the Dalitz decay domain the value $\approx0.48$ GeV$^2$ is more preferable, while in annihilation domain it comes to $0.6$ GeV$^2$ because it should reproduce experimental peak ( see the Fig \ref{fig:15}), so
\begin{equation}\label{s_8_simple}
s_8(q^2) = \begin{cases} \approx0.48 \ GeV^2, & \mbox{if } 0<|q^2|<0.3 \ GeV^2, \\ \approx0.6 \ GeV^2, & \mbox{if } |q^2|>0.3 \ GeV^2,   \end{cases}
\end{equation}

The $s_3(q^2)$ is varying from $0.6$ GeV$^2$ at $q^2\to 0$ to $0.67$ GeV$^2$ at $q^2 \to \infty$ \cite{Klopot:2011qq,Oganesian:2015ucv,Khlebtsov:2016vyf}. For the aims of this work it can be described simply by a constant $s_3=0.6$ GeV$^2$.

Let us stress out that equations \eqref{B_modified} -- \eqref{s_8_simple} represent the whole range of theoretical predictions of behaviour of the functions $B(q^2)$ and $s_{8,0}(q^2)$  on $q^2$ \cite{Khlebtsov:2020rte}.
 
Taking into account the natural assumption that the functions $B(q^2)$ as well as $s_{3,8,0}(q^2)$ should be analytical functions of $q^2$ (guaranteeing the correct analytical properties of TFF), in the next subsection we provide interpolation formulas for the EGMS16 \cite{Escribano:2015yup} mixing scheme (taken as a case study), which will describe these quantities by a smooth curves. 

\subsection{Interpolation formulas for $s_{3,8,0}$ and $B(q^2)$}
Let us consider the following functions for $s_{3,8,0}(q^2)$:
\begin{equation}\label{s8_new}
s_{3,8}(q^2) = \frac{s_{3,8}^{0.6}(1+b_{s_{3,8}}e^{-d_{s_{3,8}}(\kappa_{3,8})^2})}{1+b_{s_{3,8}}e^{-d_{s_{3,8}}(q^2)^2}},
\end{equation}
\begin{equation}\label{s0_new}
s_{0}(q^2) = \frac{s_{0}^{1.0}(1+b_{s_{0}}e^{-d_{s_{0}}(\kappa_{1})^2})}{1+b_{s_{0}}e^{-d_{s_{0}}(q^2)^2}}.
\end{equation}
We will demand that functions should reproduce poles position in the time-like annihilation domain. The values of $s_{3,8}^{0.6}$ and $s_0^{1.0}$ are fixed by positions of the experimental peaks in the time-like domain ($0.6010$ GeV$^2$ and $1.0393$ GeV$^2$, respectively), $\kappa_{3,8} = 0.6$ GeV$^2$ and $\kappa_{1} = 1.0$ GeV$^2$. The values of coefficients $b_{s_3,s_8,s_0}$ and $d_{s_{3},s_{8},s_{0}}$ we will estimate from the experimental data.

Turn on to the function $B(q^2)$. We will demand that it must reproduce the value $B(0)$. This quantity is evaluated from $P\rightarrow\gamma\gamma$ decay width, which was measured with very high accuracy with a few $\%$  error. Consider the following parametrisation 
\begin{align}
B(q^2) &= \frac{a_{B}}{1+(\frac{a_{B}}{B(0)-c_1} - 1)e^{-d_{B}(q^2)^2}} + B_{max}e^{-\frac{(q^2 -\mu)^2}{2\sigma^2}},  \\
c_1 &= B_{max}e^{-\frac{(0.0 -\mu)^2}{2\sigma^2}},
\end{align}\label{b_new}
with resonance parameters $B_{max} = -1.1, \sigma = 0.3$ GeV$^2$, $\mu = 1.0$ GeV$^2$. The value $B(0)$ at $q^2 = 0$ is taken from the Table.\ref{table2}. The values of coefficients $a_{B}$ and $d_{B}$ will be estimated using the experimental data similarly to the case of $s_{3,8,0}(q^2)$ functions.

Note that the proposed form of interpolation formulas for $s_{3,8,0}(q^2)$ and $B(q^2)$ is not unique. At the same time, any other interpolation formulas for $s_{3,8,0}(q^2)$ and $B(q^2)$ must satisfy the conditions listed above, which represent real physical meaning of these quantities.

In order to estimate unknown coefficients in \eqref{s8_new}, \eqref{s0_new} and \eqref{b_new} we are going to take combined space-like and Dalitz domains $\eta+\eta'$ meson dataset \cite{Acciarri:1998, Behrend:1990sr, Gronberg:1997fj, BABAR:2011ad, Adlarson:2016hpp, Arnaldi:2016pzu, Ablikim:2015wnx}. The established values of coefficients for the EGMS16 \cite{Escribano:2015yup} mixing scheme are listed in the Table.\ref{table3}, while in the Table.\ref{table4}  the obtained $\chi^2$ and slopes values are shown.
\begin{table}[H]
\caption{The values of $a,b,d$ coefficients from fitting combined space-like and Dalitz domains $\eta+\eta'$ meson data. The $1\sigma$ error is shown in brackets.}\label{table3}
\resizebox{1.0\textwidth}{!}{\begin{tabular}{c|c|c|c|c|c|c|c}
$b_{s_3}$ & $d_{s_3}$, GeV$^{-4}$  & $b_{s_8}$ & $d_{s_8}$, GeV$^{-4}$ & $b_{s_0}$ & $d_{s_0}$, GeV$^{-4}$ & $a_{B}$ & $d_{B}$, GeV$^{-4}$  \\ \hline
0.101 & 82 & 0.13(0.1) & 15.45(8.22) & 1.46(0.21) & 2.25(0.53) & -0.364(0.042) & 1.99(0.76)
\end{tabular}}
\end{table}
\begin{table}[H]
\caption{The $\chi^2$ values in space-like and Dalitz domains, with the number of experimental points listed in brackets. In the last column the slopes of TFFs are shown.}\label{table4}
\begin{tabular}{c|c|c|c}
Meson & Space-like & Dalitz & Slope  \\ \hline
$\eta$ & 75.0(34) & 13.9(48) & 1.97 \\ 
$\eta'$ & 48.6(50) & 12.4(8) & 2.06 
\end{tabular}
\end{table}
The corresponding graphs of $s_{3,8,0}(q^2)$ \eqref{s8_new},\eqref{s0_new} and $B(q^2)$ \eqref{b_new} with obtained coefficients listed in Table\ref{table3} are shown in Fig.\ref{fig:18} and  \ref{fig:19} respectively.
\begin{figure}[H]
 \begin{floatrow}
   \ffigbox{\caption{The functions $s_{3,8,0}(q^2)$ \eqref{s8_new},\eqref{s0_new}}. \label{fig:18}}%
           {\includegraphics[width=0.5\textwidth]{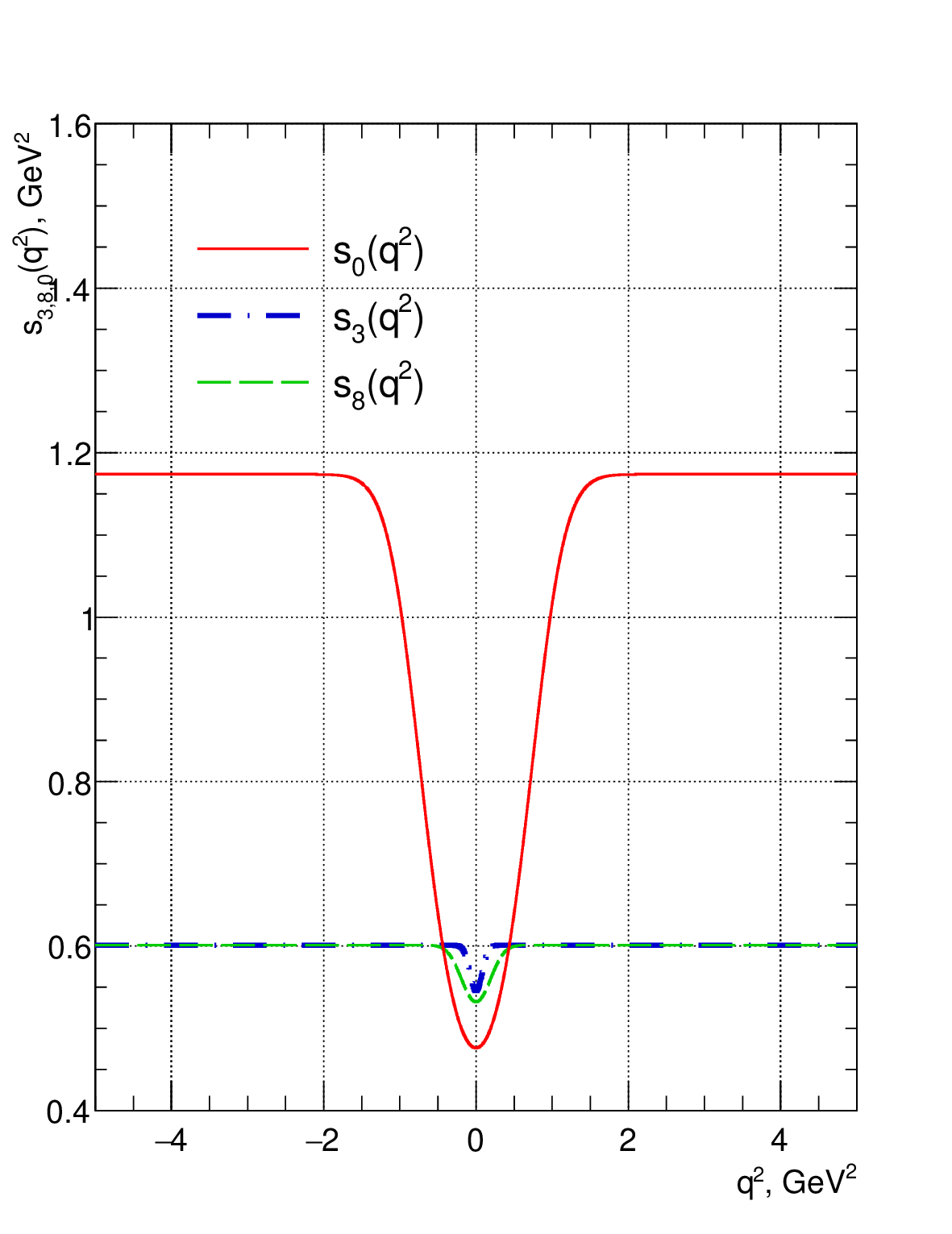}}
   \ffigbox{\caption{The functions $B(q^2)$ \eqref{b_new}.}\label{fig:19}}%
           {\includegraphics[width=0.5\textwidth]{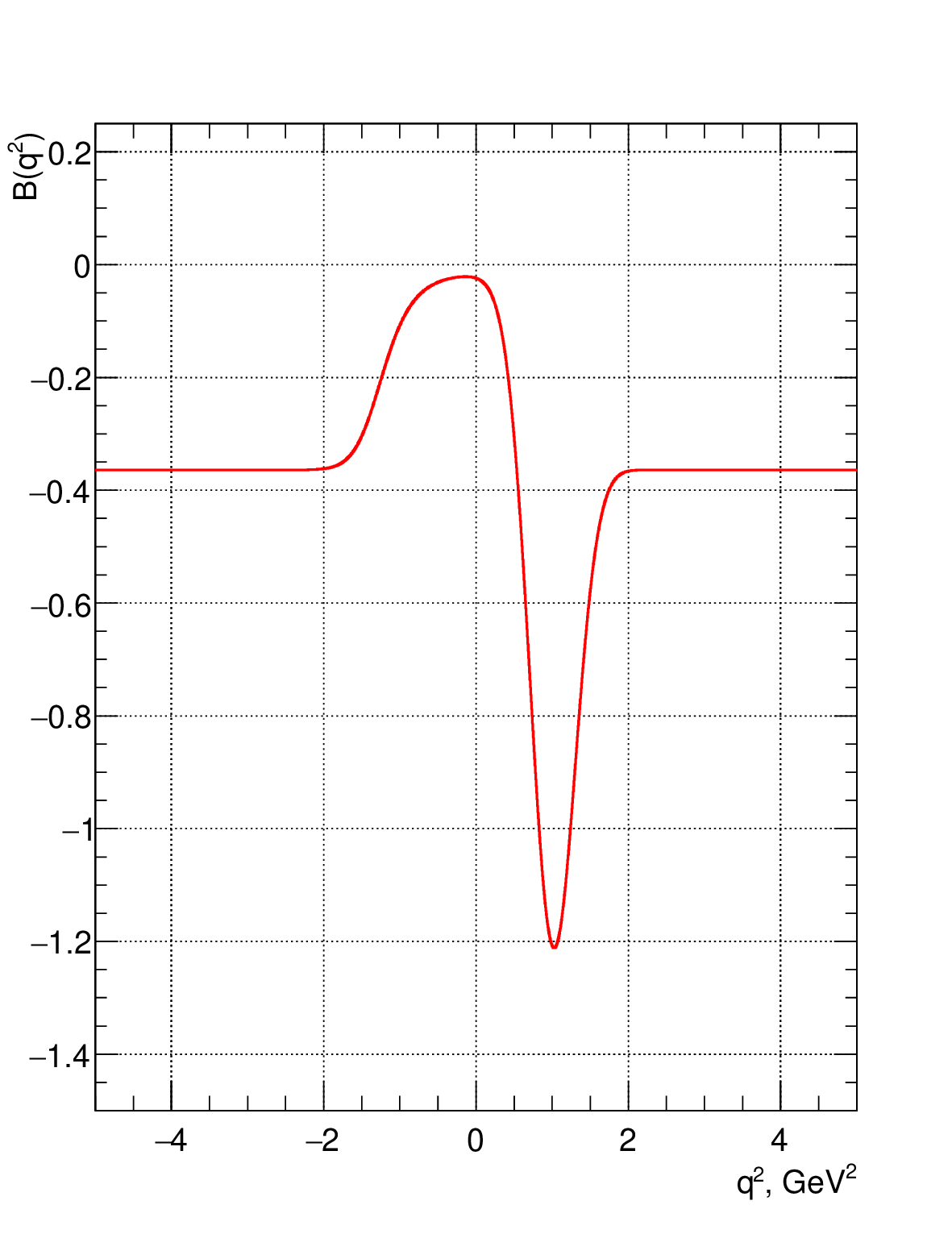}}         
  \end{floatrow}
\end{figure}

The $\pi^0,\eta,\eta'$ TFFs graphs with obtained coefficients( in the Table.\ref{table3}) are shown in Fig.\ref{fig:20}-\ref{fig:25}
\begin{figure}[H]
 \begin{floatrow}
   \ffigbox{\caption{The $\eta\rightarrow \gamma\gamma^*$ TFF \eqref{solution_generic} in the space-like region  with the interpolation formulas for $s_{3,8,0}(q^2)$ \eqref{s8_new},\eqref{s0_new} and $B(q^2)$ \eqref{b_new}.  The insert shows a low $Q^2$ region. The experimental data are taken from \cite{Behrend:1990sr, Gronberg:1997fj, BABAR:2011ad}}. \label{fig:20}}%
           {\includegraphics[width=0.5\textwidth]{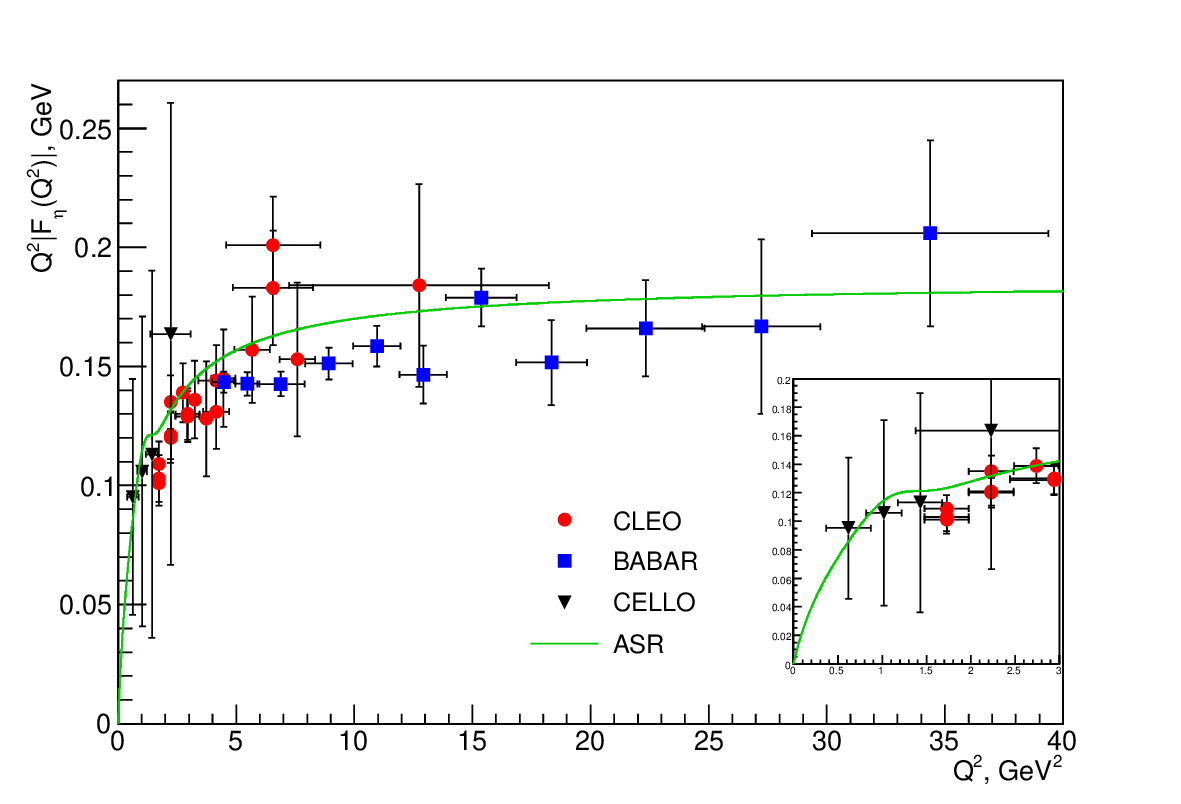}}
   \ffigbox{\caption{The $\eta'\rightarrow \gamma\gamma^*$ TFF \eqref{solution_generic} in the space-like region with the interpolation formulas for $s_{3,8,0}(q^2)$ \eqref{s8_new},\eqref{s0_new} and $B(q^2)$ \eqref{b_new}.  The insert shows a low $Q^2$ region. The experimental data  are taken from  \cite{Acciarri:1998, Behrend:1990sr, Gronberg:1997fj, BABAR:2011ad}. For comparison the one-pole parametrisation for TFF from BESIII collaboration \cite{Ablikim:2015wnx}( equation (4)) is shown by dot-dashed curve. }\label{fig:21}}%
           {\includegraphics[width=0.5\textwidth]{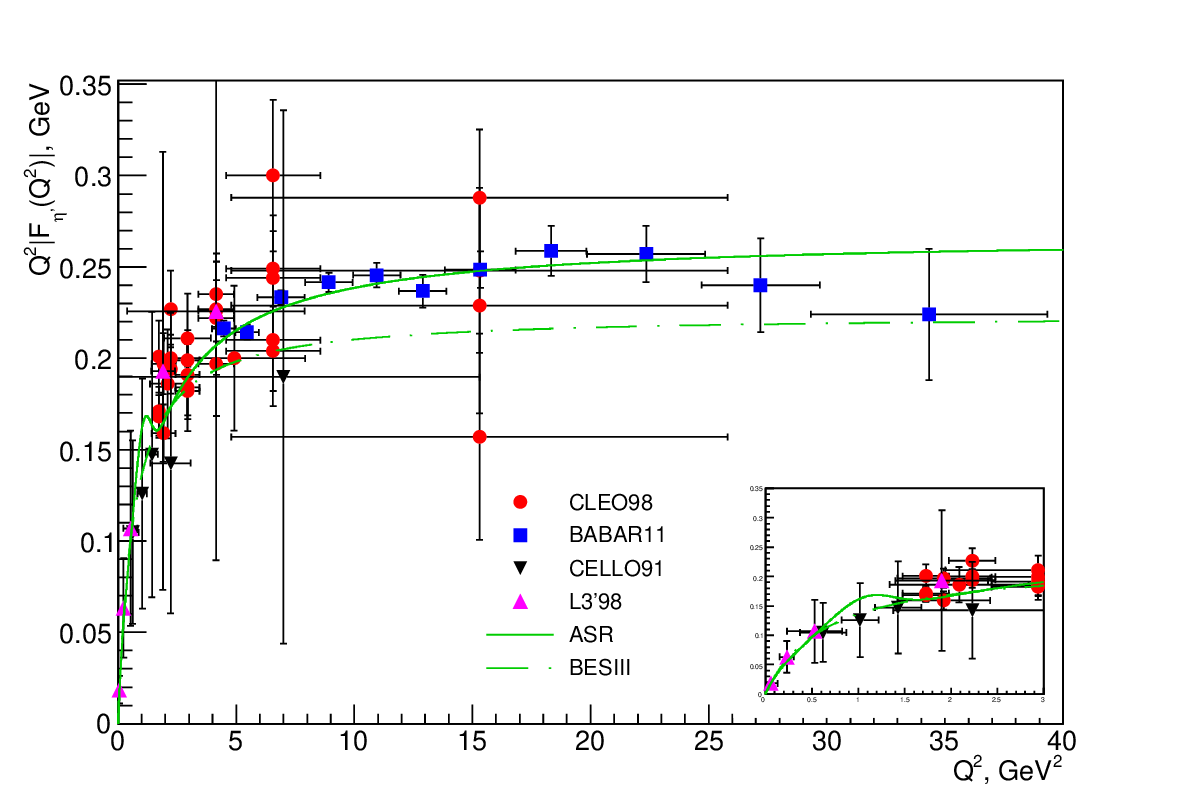}}         
  \end{floatrow}
\end{figure}

\begin{figure}[H]
	\begin{floatrow}
		\ffigbox{\caption{The $\eta$ TFFs \eqref{solution_generic_time} in the Dalitz domain with the interpolation formulas for $s_{3,8,0}(q^2)$ \eqref{s8_new},\eqref{s0_new} and $B(q^2)$ \eqref{b_new}. The experimental data are by A2  \cite{Adlarson:2016hpp} and NA60 \cite{Arnaldi:2016pzu} collaborations.} \label{fig:22}}%
		{\includegraphics[width=0.5\textwidth]{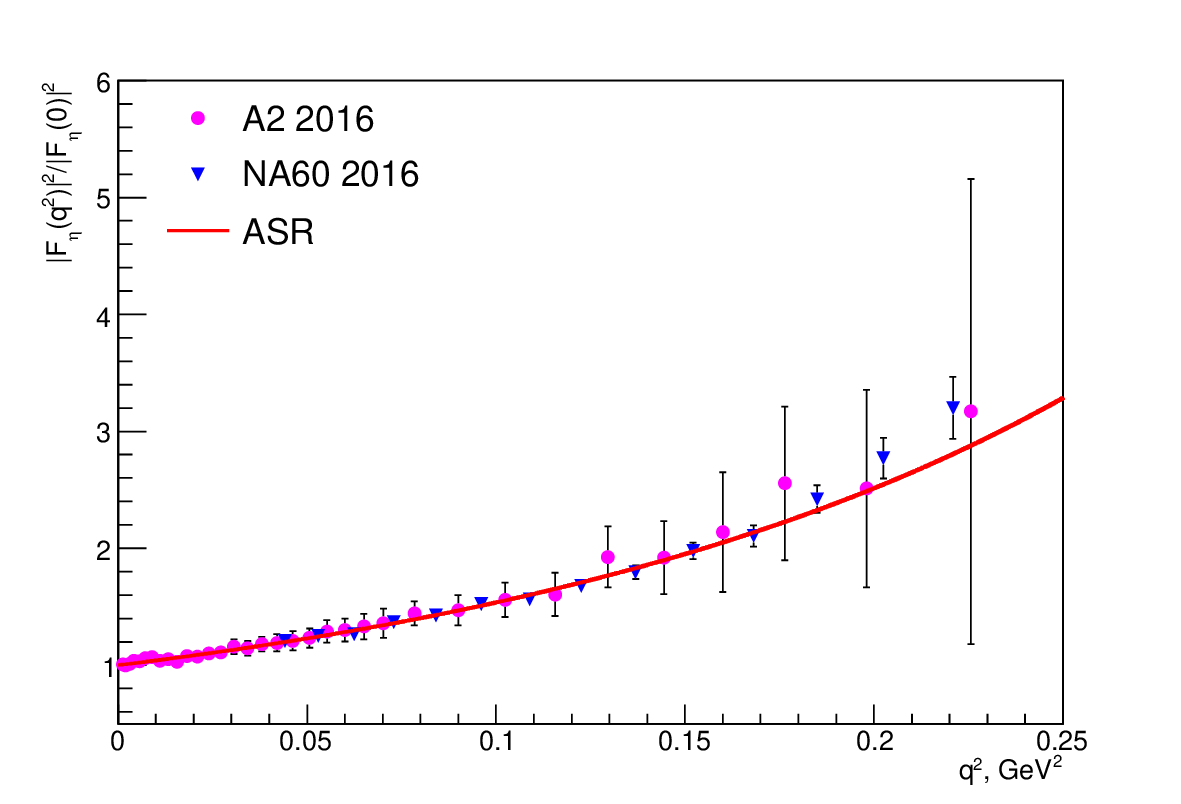}}
		\ffigbox{\caption{The $\eta'$ TFFs \eqref{solution_generic_time} in the Dalitz domain with the interpolation formulas for $s_{3,8,0}(q^2)$ \eqref{s8_new},\eqref{s0_new} and $B(q^2)$ \eqref{b_new}. The experimental data are by BESIII collaboration \cite{Ablikim:2015wnx}. }\label{fig:23}}%
		{\includegraphics[width=0.5\textwidth]{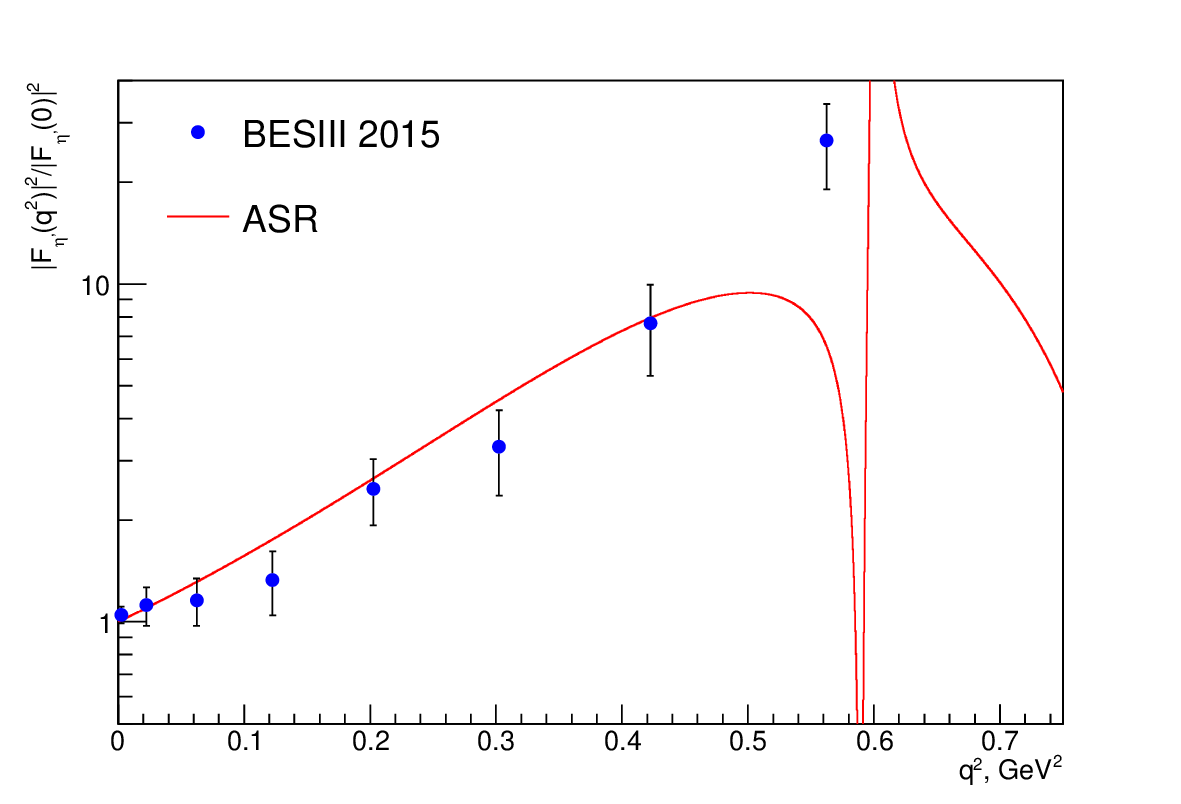}}         
	\end{floatrow}
\end{figure}

\begin{figure}[H]
	\begin{floatrow}
		\ffigbox{\caption{$\pi^0$ TFF \eqref{solution_generic_time} with the interpolation formulas for $s_{3,8,0}(q^2)$ \eqref{s8_new},\eqref{s0_new} and $B(q^2)$ \eqref{b_new} compared with the experimental data \cite{Adlarson:2016ykr,Achasov:2016bfr,Akhmetshin:2004gw,Achasov:2018ujw}.  The insert shows the Dalitz decay domain.} \label{fig:24}}%
		{\includegraphics[width=0.5\textwidth]{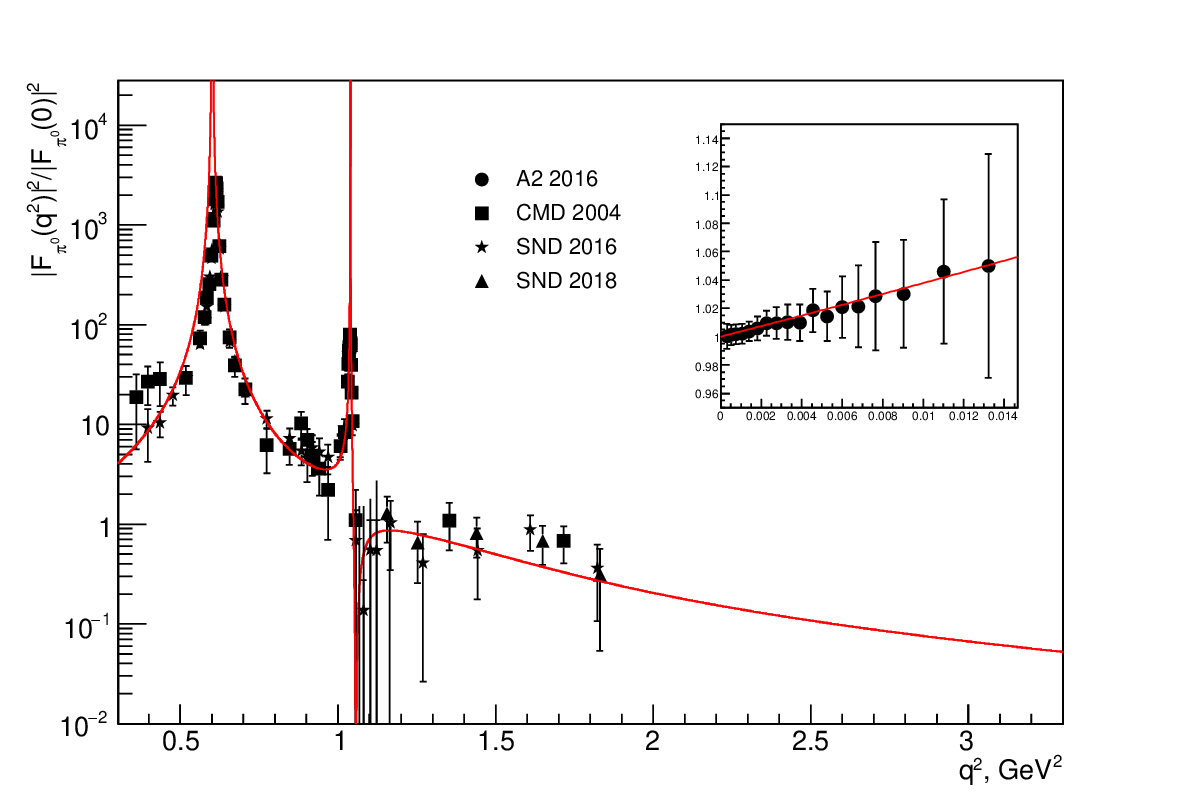}}
		\ffigbox{\caption{The $\eta$ TFF \eqref{solution_generic_time} with the interpolation formulas for $s_{3,8,0}(q^2)$ \eqref{s8_new},\eqref{s0_new} and $B(q^2)$ \eqref{b_new} compared with the experimental data \cite{Achasov:2007kw,Achasov:2013eli,Akhmetshin:2004gw}. }\label{fig:25}}%
		{\includegraphics[width=0.5\textwidth]{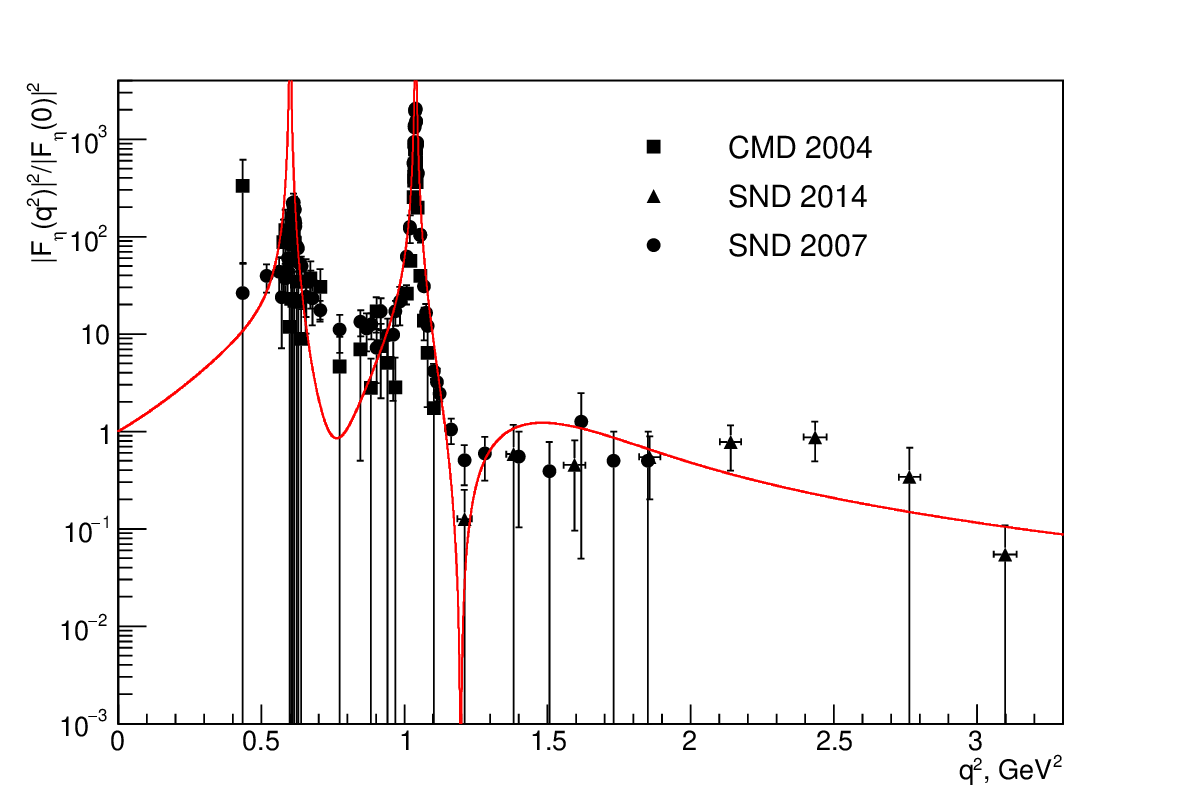}}         
	\end{floatrow}
\end{figure}

At high energy region  $q^2 = 112$ GeV$^2$ \cite{Aubert:2006cy} the experimental values are $q^2|F_{\eta}(q^2)| = 0.229(0.031)$ and $q^2|F_{\eta'}(q^2)| = 0.251(0.02)$. In our approach the equation \eqref{solution_generic_time} with the interpolation formulas for $s_{3,8,0}(q^2)$ \eqref{s8_new},\eqref{s0_new} and $B(q^2)$ \eqref{b_new} gives: $q^2|F_{\eta}(q^2)| = 0.184$ and $q^2|F_{\eta'}(q^2)| = 0.264$.  

Thus the proposed interpolation formulas for $s_{3,8,0}(q^2)$ \eqref{s8_new}, \eqref{s0_new} and $B(q^2)$ \eqref{b_new} with obtained coefficients listed in the Table \ref{table3} provide good description of the $\pi^0$, $\eta$ and $\eta'$ TFFs both in space- and time-like regions. Let us note that in the case of other mixed schemes interpolation formulas for $s_{3,8,0}(q^2)$ and $B(q^2)$ can be different than \eqref{s8_new}, \eqref{s0_new} and \eqref{b_new}. However as can be seen in Appendix \ref{appendix_a} the decay constants in FKS98 \cite{Feldmann:1998vh}, EF05 \cite{Escribano:2005qq} and KOT12 \cite{Klopot:2012hd} mixing schemes do not  differ strongly from the EGMS16 \cite{Escribano:2015yup}, so the curves for $s_{3,8,0}(q^2)$ and $B(q^2)$ obtained with other possible interpolation formulas for these mixing schemes will be similar to EGMS16 \cite{Escribano:2015yup} one.

\section{$P\rightarrow Z^{0}\gamma$ transitions }
Now we are going to consider the $\pi^0$, $\eta$ and $\eta'$ transition form factors  with one photon and one $Z^0$-boson. 

Let us briefly recall the neutral weak current properties. Following \cite{Okun:1982ap} in general the neutral weak current can be written as a sum of 12 terms:
\begin{equation}\label{neutral}
j_{\mu}^{0}=\sum_i(g^i_{L}\bar{\psi_i}O^{L}_{\mu}\psi_i + g^i_{R}\bar{\psi_i}O^{R}_{\mu}\psi_i),
\end{equation} 
where
\begin{align}\label{who}
i = \nu_e, \ \nu_{\mu}, \ \nu_{\tau}, \ e, \ \mu, \ \tau, \ u, \ c, \ t, \ d, \ s,\ b, \\
O_{\mu}^{L} = \gamma_{\mu}(1 + \gamma_5), \ O_{\mu}^{R} = \gamma_{\mu}(1 - \gamma_5),   
\end{align}
$g^i_{L}$ and $g^i_{R}$ are the numerical coefficients corresponding to weak charges 
\begin{align}
g^i_{L} & =  \frac{1}{2}, \ \ \ g^i_{R}  =  0 \ for \ \nu_e, \ \nu_{\mu}, \ \nu_{\tau}, \\
g^i_{L} & =  -\frac{1}{2}+\xi, \ g^i_{R}  =  +\xi \ for \ \ e, \ \mu, \ \tau, \\
g^i_{L} & =  \frac{1}{2}-\frac{2}{3}\xi, \ g^i_{R}  =  -\frac{2}{3}\xi \ for \ u, \ c, \ t, \\
g^i_{L} & =  -\frac{1}{2}+\frac{1}{3}\xi, \ g^i_{R}  =  \frac{1}{3}\xi \ for \ d, \ s,\ b, 
\end{align}
$\xi=sin^2\theta_{W}$, where $\theta_{W}$ is the Weinberg angle( $\xi\approx 0.23$).

As seen from \eqref{neutral} and \eqref{who} the neutral weak current has both vector(V) and axial(A) parts. Let us rewrite the equations for weak currents separating V and A parts.
\begin{equation}\label{neutral_VA}
j_{\mu}^{0}=\sum_i [ (g^i_{L}+g^i_{R})\bar{\psi_i}\gamma_{\mu}\psi_i + (g^i_{L}-g^i_{R})\bar{\psi_i}\gamma_{\mu}\gamma_5\psi_i ].
\end{equation}
In the case  under consideration we already have vector EM current from photon and one axial current. The AVA diagram can be neglected, so only vector part of neutral currents remains in \eqref{neutral_VA}. Thus the triangle graph amplitude for the processes involving  $Z^0$-boson and photon will be similar to the 2 photons case. Technically this diagram  differs only by the charge factor at one vertex. The corresponding 3-point correlation function contains the axial current $J_{\alpha5}$ with momentum $p = k + q$, EM vector current $J_{\nu}=\sum_i {e_i\bar{\psi_i}\gamma_{\nu}\psi_i}, \ i=u,d,s$ with momenta $k$(photon) and vector part of the  neutral weak current \eqref{neutral_VA} with momenta $q$($Z^0$-boson)

The transition of isovector and octet axial currents will have the matrix element $\langle 0 |F\tilde{F} |Z^0\gamma \rangle$ stemmed from the Abelian (electromagnetic) anomaly, while the singlet axial current will have an additional matrix element $\langle 0 |G\tilde{G} |Z^0\gamma \rangle$ stemmed from the Non-Abelian (strong) anomaly:
	\begin{equation}\label{an-0_weak}
	\partial^\mu J_{\mu 5}^{(0)} =\frac{2i}{\sqrt 3}\sum_i{m_i \bar{\psi_i} \gamma_5 \psi_i}+\frac{v}{\sqrt{3}\pi^2}N_c  F\widetilde{F} + \frac{n_f\alpha_s}{4\pi\sqrt{3}}  G\widetilde{G},
	\end{equation}
	\begin{equation}\label{an-3_weak}
	\partial^\mu J_{\mu 5}^{(3)} =\frac{2i}{\sqrt2}\sum_i{m_i \bar{\psi_i} \gamma_5 \lambda^3 \psi_i} 
	+\frac{v}{\sqrt{2}\pi^2}N_c  F\widetilde{F},
	\end{equation}
	\begin{equation}\label{an-8_weak}
	\partial^\mu J_{\mu 5}^{(8)} =\frac{2i}{\sqrt2}\sum_i{m_i \bar{\psi_i} \gamma_5 \lambda^8 \psi_i} 
	+\frac{v}{\sqrt{6}\pi^2}N_c  F\widetilde{F},
	\end{equation}
where $v=\sum_ke_kx_k$ and $k=u,d,s$-quarks. The $e_k$ factors denote EM quark charges ( in the electron charge $e$ units) and $x_k$ are the $Z^0$-boson coupling "charges" $x_u=\bar{g}(\frac{1}{2}-\frac{4}{3}\xi)$ and $x_{d,s}=\bar{g}(-\frac{1}{2}+\frac{2}{3}\xi)$ for $u-$ and $d,s-$quarks, respectively.

The $Z^0$-boson coupling "charges" for $u-$ and $d,s-$quarks are different and so the Feynman diagrams for the matrix element $\langle 0 |F\tilde{F} |Z^0\gamma \rangle$ now should be changed to two diagrams depicted in the Fig.\ref{fig:26} and \ref{fig:27}.
\begin{figure}[H]
\begin{floatrow}
			\ffigbox{\caption{$A^{Z^0}_{QED}$ for u-quark with $Z^0$ boson vertex.} \label{fig:26}}%
			{\includegraphics[width=0.525\textwidth]{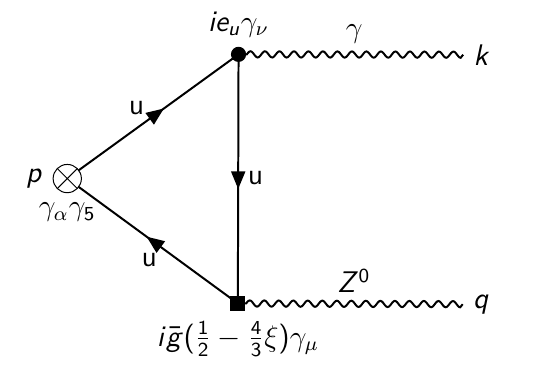}}
			\ffigbox{\caption{$A^{Z^0}_{QED}$ for d,s-quarks with $Z^0$ boson vertex.}\label{fig:27}}%
			{\includegraphics[width=0.5\textwidth]{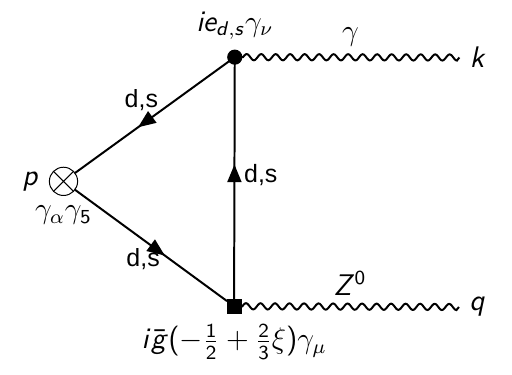}}  \end{floatrow}
\end{figure}
For the charge factors we got for $C^{(0)}_{Z^0}$
\begin{align}\label{ch_fctrs_Zgamma}
C^{(0)}_{Z^0} &= \frac{1}{\sqrt{3}}\left(e_u\bar{g}(\frac{1}{2} - \frac{4}{3}\xi) + e_d\bar{g}(-\frac{1}{2} + \frac{2}{3}\xi) + e_s\bar{g}(-\frac{1}{2} + \frac{2}{3}\xi)\right) = \\
& = \frac{e\bar{g}}{\sqrt{3}}\left(\frac{2}{3}(\frac{1}{2} - \frac{4}{3}\xi) + (-\frac{1}{3})(-\frac{1}{2} + \frac{2}{3}\xi) + (-\frac{1}{3})(-\frac{1}{2} + \frac{2}{3}\xi)\right)
= \frac{2e\bar{g}}{3\sqrt{3}}(1-2\xi). 
\end{align}
It can be rewritten by using charge factor for the case of 2 photons $C^{(0)}$ \eqref{ch_fctrs_2gamma} as
\begin{equation}
C^{(0)}_{Z^0} = C^{(0)}\frac{\bar{g}}{e}(1-2\xi).
\end{equation}
Similarly one can obtain such relations for isovector and octet currents
\begin{equation}
C^{(3)}_{Z^0} = C^{(3)}\frac{\bar{g}}{e}(\frac{1}{2}-2\xi),
\end{equation}
\begin{equation}
C^{(8)}_{Z^0} = C^{(8)}\frac{\bar{g}}{e}(\frac{1}{2}-2\xi).
\end{equation}

As seen in \eqref{an-0_weak} and \eqref{an-3_weak}, \eqref{an-8_weak} the term $F\tilde{F}$ differs from the one in equations \eqref{an-0} and \eqref{an-38} by a charge factor constant. The interaction constant at $Z^0$-boson vertex doesn't affect the kinematics, one can use the vector-vector-axial (VVA) amplitude decomposition from \eqref{VVA}.  It means that the equations of the ASR for isovector ($a=3$) and octet ($a=8$) currents with one real photon ($k^2=0$) and $Z^0$-boson ($Q^2=-q^2 \geq 0$) will be similar to \eqref{asr38} with the only difference in the charge factor coefficients (in what follows we put $m_u=m_d=m_s=0$)
\begin{equation}\label{asr38_weak}
	\frac{1}{\pi}\int_{0}^{\infty}A^{(3,8)}_{Z^0}(s,Q^2)ds = \frac{C^{(3,8)}_{Z^0}N_c}{2\pi^2},
\end{equation}
where the spectral density function is defined as $A^{(3,8)}_{Z^0} \equiv \frac{1}{2}Im(F_3-F_6)$ and is determined from the VVA amplitude decomposition \eqref{VVA}. The one-loop approximation for the spectral densities of the isovector and octet currents $A_3^{(3,8)}(s,Q^2)$ is 
\begin{align} \label{a3_weak}
A_{Z^0}^{(3,8)}(s, Q^2)=\frac{C^{(3,8)}_{Z^0}N_c}{2\pi}\frac{Q^2}{(s+Q^2)^2},
\end{align}
and the ASRs for the hadron contributions are
\begin{equation} \label{asr38aa_weak}
	\Sigma f_P^{(3,8)}F_{PZ^0\gamma}(Q^2) = \frac{C^{(3,8)}_{Z^0}N_c}{2\pi^2}\frac{s_{3,8}}{s_{3,8}+Q^2}.
\end{equation}

For the singlet current alongside the matrix element $\langle 0 |F\tilde{F} |Z^0\gamma \rangle$  the non-perturbative matrix element $\langle 0 |G\tilde{G}|\gamma Z^{0} \rangle$ corresponding to strong anomaly term in \eqref{an-0_weak} will contribute:
\begin{equation} \label{N_weak}
	\langle 0 | \frac{\sqrt{3}\alpha_s}{4\pi} G\tilde{G}|\gamma Z^{0} \rangle = C^{(0)}_{Z^0}N_c N_{Z^0}(p^2, k^2,q^2) \epsilon^{\mu\nu\rho\sigma}k_{\mu}q_{\nu}\epsilon_{\rho}^{(k)}\epsilon_{\sigma}^{(q)}.
	\end{equation}
This matrix element can be schematically represented by a diagram shown in the Fig.\ref{fig:200}. The shaded circle denotes all possible perturbative and non-perturbative transitions of two gluons to $Z^0$-boson and photon. The black circle corresponds to EM vertex with $e_{u,d,s}$ EM quark charges (in the units of electron charge $e$). The black square represents vertex corresponding to outgoing $Z^0$-boson, where $Z^0$-boson coupling "charges" are $x_u=\bar{g}(\frac{1}{2}-\frac{4}{3}\xi)$ and $x_{d,s}=\bar{g}(-\frac{1}{2}+\frac{2}{3}\xi)$ for $u-$ and $d,s-$quarks, respectively. 
\begin{figure}[H]
\caption{The matrix element $\langle 0 | \frac{\sqrt{3}\alpha_s}{4\pi} G\tilde{G}|Z^0\gamma \rangle$ schematic representation.}
\includegraphics[width=0.5\textwidth]{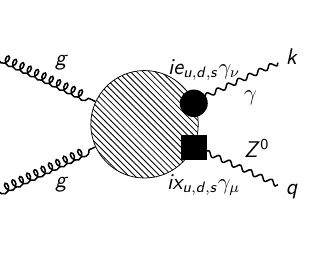}\label{fig:200}
\end{figure}
For this matrix element one need to  sum over all $u,d,s$-quark charges products, so at the leading order in $\frac{1}{m_{W}^2}$ it has the same charge factor coefficient $C^{(0)}_{Z^0}$ as the matrix element $\langle 0 |F\tilde{F} |Z^0\gamma \rangle$. The factor $N_c$ is indicated for convenience.

The corresponding non-Abelian contribution in the dispersive form requires a subtraction, so the singlet current ASR in the considered kinematics $N(p^2, k^2=0, q^2=-Q^2)\equiv N(p^2, Q^2)$ can be written as  
\begin{equation}\label{asr0_weak}
\frac{1}{\pi}\int_{0}^{\infty}A^{(0)}_{Z^0}(s,Q^{2})ds = \frac{C^{(0)}_{Z^0}N_c}{2\pi^2} + C^{(0)}_{Z^0}N_c\left(N_{Z^0}(0,Q^2)-\frac{1}{\pi}\int_{0}^{\infty}Im R_{Z^0}(s,Q^2)ds\right),
\end{equation}
where 
\begin{equation}\label{N_sub_weak}
	R_{Z^0}(p^2,Q^2)=\frac{1}{p^2}(N_{Z^0}(p^2,Q^2)-N_{Z^0}(0,Q^2)).
\end{equation}

In order to single out electromagnetic contribution one can  split the spectral density of singlet current into two parts,
\begin{equation}\label{split_weak}
	A^{(0)} = A_{QED}^{Z^0}+A_{QCD}^{Z^0}.
\end{equation}
The first part in \eqref{split_weak} $A_{QED}^{Z^0}$ represents the contribution of QED diagrams, whose lowest one-loop part (Fig. \ref{fig:26},\ref{fig:27}) is given by the expression  \eqref{a3_weak} with an appropriate charge factor $C^{(0)}_{Z^0}$.

The second part $A_{QCD}^{Z^0}$ is the contribution of diagrams (Fig.~\ref{fig:55}) with virtual gluons coupled to photon and $Z^0$-boson through all possible perturbative and non-perturbative interactions.
\begin{figure}[H]
\caption{$A_{QCD}^{Z^{0}}$ for u,d,s-quarks with $Z^0$ boson vertex. The black circle corresponds to EM vertex with $e_{u,d,s}$ EM quark charges (in the units of electron charge $e$). The black square represents vertex corresponding to outgoing $Z^0$-boson, where $Z^0$-boson coupling "charges" are $x_u=\bar{g}(\frac{1}{2}-\frac{4}{3}\xi)$ and $x_{d,s}=\bar{g}(-\frac{1}{2}+\frac{2}{3}\xi)$ for $u-$ and $d,s-$quarks, respectively.} \label{fig:55}%
\includegraphics[width=0.75\textwidth]{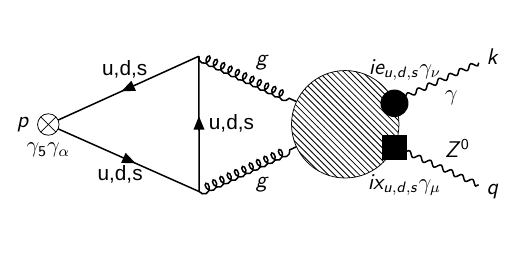}
\end{figure}

Thus one can rewrite the ASR \eqref{asr0_weak} as
\begin{equation}\label{qedplusqcd_weak}
	\Sigma f_P^{(0)}F_{PZ^0\gamma}(Q^2) = \frac{N_cC^{(0)}_{Z^0}}{2\pi^2}\frac{s_0}{s_0+Q^2} - C^{(0)}_{Z^0}N_c\left(\frac{1}{\pi}\int_{s_0}^{\infty} A_{QCD}^{Z^0}ds + N_{Z^0}(0,Q^2) - \frac{1}{\pi}\int_{0}^{\infty} ImR_{Z^0}(s,Q^{2})ds\right).
\end{equation} 
	The first and the last three terms in Eq. (\ref{qedplusqcd_weak}) represent the electromagnetic and the strong anomaly contributions to the ASR, respectively.  Let us introduce the function that represents the strong and electromagnetic anomalies contributions ratio:
\begin{equation}\label{gluon_anom_contrib1_weak}
B_{Z^0}(Q^2, s_0) = \frac{2\pi^2}{N_cC^{(0)}_{Z^0}}\frac{s_0+Q^2}{s_0}  \left[C^{(0)}_{Z^0}N_c\left(N_{Z^0}(0,Q^2) - \frac{1}{\pi}\int_{0}^{\infty} ImR_{Z^0}(s,Q^{2})ds - \frac{1}{\pi}\int_{s_0}^{\infty} A_{QCD}^{Z^0}(s,Q^2)ds\right) \right].
\end{equation} 
As the integral of $A_{QCD}^{Z^0}$ is suppressed as $\alpha^2_s$ at $s_0 \geq 1.0$ GeV$^2$, the function $B_{Z^0}(Q^2,s_0)$ is predominantly determined by the first two terms. It reflects the properties of the non-perturbative matrix element  $\langle 0 |G\tilde{G}|\gamma Z^{0} \rangle$. 

And finally the ASR for the singlet current \eqref{qedplusqcd_weak} in terms of the function $B_{Z^0}(Q^2,s_0)$ \eqref{gluon_anom_contrib1_weak} reads:
\begin{equation}\label{mix0_weak}
\Sigma f_P^{(0)}F_{PZ^0\gamma}(Q^2) = \frac{N_cC_{Z^0}^{(0)}}{2\pi^2}\frac{s_0}{s_0+Q^2} \left[1+B_{Z^0}(Q^2, s_0) \right].
\end{equation}

The function $B_{Z^0}(Q^2, s_0)$ is unknown and cannot be calculated analytically due to non-pertubative origin of the corresponding matrix element. As it was pointed out in the previous section, it is the  strong and EM contributions ratio. Therefore, as  the charge factor coefficients of the corresponding matrix elements appear to be the same, the function $B_{Z^0}(Q^2, s_0)$ does not depend on these coefficients.

Strictly speaking, functions $B_{Z^0}(Q^2, s_0)$ in \eqref{gluon_anom_contrib1_weak} and $B(Q^2, s_0)$ are different \eqref{gluon_anom_contrib1}. But the matrix element $\langle 0 |G\tilde{G}|\gamma Z^{0} \rangle$ up to $\frac{1}{m_{W}^2}$ corrections differs from the matrix element $\langle 0 |G\tilde{G}|\gamma \gamma \rangle$ only by a charge factor ($N=N_Z^0$).  For the matrix elements corresponding to EM anomaly contributions it is the same difference. Thus to the first order of weak corrections the functions $B_{Z^0}(Q^2, s_0)$ and $B(Q^2, s_0)$ are equal. In the previous section it was shown that in paper \cite{Khlebtsov:2020rte} the $B(Q^2, s_0)$ behaviour was established from $P\rightarrow\gamma\gamma^{(*)}$ TFFs and it can be described by the function shown in the Fig.\ref{fig:17}.

Thus solving system of equations of the  ASRs for the isovector, octet (\ref{asr38aa_weak}) and singlet (\ref{mix0_weak}) currents
\begin{equation}\label{system_Z}
	\left(
	\begin{matrix}
	f_{\pi^0}^{(3)} & f_{\eta}^{(3)} & f_{\eta'}^{(3)} \\
	f_{\pi^0}^{(8)} & f_{\eta}^{(8)} & f_{\eta'}^{(8)} \\
	f_{\pi^0}^{(0)} & f_{\eta}^{(0)} & f_{\eta'}^{(0)} 
	\end{matrix}
	\right)
	\left(
	\begin{matrix}
	F_{\pi^0}(Q^2) \\
	F_{\eta}(Q^2)  \\
	F_{\eta'}(Q^2)
	\end{matrix}
	\right)=
	\left(
	\begin{matrix}
	\frac{N_cC^{(3)}_{Z^0}}{2\pi^2}\frac{s_3}{s_3+Q^2} \\
	\frac{N_cC^{(8)}_{Z^0}}{2\pi^2}\frac{s_8}{s_8+Q^2}  \\
	\frac{N_cC^{(0)}_{Z^0}}{2\pi^2}\frac{s_0(1+B(Q^2,s_0))}{s_0+Q^2}
	\end{matrix}
	\right),
	\end{equation}
one obtains expressions for the transition form factors:
	\begin{equation}\label{solution_generic_Z}
	F_{P\gamma Z^{0}}(Q^2) = \bar{\alpha}_{P} \frac{s_3}{s_3+Q^2} + \bar{\beta}_{P} \frac{s_8}{s_8+Q^2} + \bar{\gamma}_{P} \frac{s_0}{s_0+Q^2}[1+B_{Z^0}(Q^2,s_0)],
\end{equation}
	where $P = \pi^0, \eta, \eta'$. The coefficients $\alpha_P$, $\beta_P$, $\gamma_P$ are expressed in terms of the decay constants $f_P^{(i)}$ listed in the Table \ref{table5},  $\Delta$ is the determinant of the decay constant matrix in (\ref{system_Z}):
	\begin{align}
	\small &\bar{\alpha}_{\pi^0}= \frac{N_cC^{(3)}_{Z^0}}{2\pi^2\Delta}(f^{(8)}_{\eta}f^{(0)}_{\eta'} - f^{(0)}_{\eta}f^{(8)}_{\eta'}),&\bar{\beta}_{\pi^0}= \frac{N_cC^{(8)}_{Z^0}}{2\pi^2\Delta}(f^{(0)}_{\eta}f^{(3)}_{\eta'} - f^{(3)}_{\eta}f^{(0)}_{\eta'}),&\bar{\gamma}_{\pi^0}= \frac{N_cC^{(0)}_{Z^0}}{2\pi^2\Delta}(f^{(3)}_{\eta}f^{(8)}_{\eta'} - f^{(8)}_{\eta}f^{(3)}_{\eta'}), \label{coef_pi_Z}\\ \label{coef_eta_Z}
	&\bar{\alpha}_{\eta}= \frac{N_cC^{(3)}_{Z^0}}{2\pi^2\Delta}(f^{(0)}_{\pi^0}f^{(8)}_{\eta'} - f^{(8)}_{\pi^0}f^{(0)}_{\eta'}),&\bar{\beta}_{\eta}= \frac{N_cC^{(8)}_{Z^0}}{2\pi^2\Delta}(f^{(3)}_{\pi^0}f^{(0)}_{\eta'} - f^{(0)}_{\pi^0}f^{(3)}_{\eta'})\;,&\bar{\gamma}_{\eta}= \frac{N_cC^{(0)}_{Z^0}}{2\pi^2\Delta}(f^{(8)}_{\pi^0}f^{(3)}_{\eta'} - f^{(3)}_{\pi^0}f^{(8)}_{\eta'}),\\ 
	&\bar{\alpha}_{\eta'}= \frac{N_cC^{(3)}_{Z^0}}{2\pi^2\Delta}(f^{(8)}_{\pi^0}f^{(0)}_{\eta} - f^{(0)}_{\pi^0}f^{(8)}_{\eta}),&\bar{\beta}_{\eta'}= \frac{N_cC^{(8)}_{Z^0}}{2\pi^2\Delta}(f^{(0)}_{\pi^0}f^{(3)}_{\eta} - f^{(3)}_{\pi^0}f^{(0)}_{\eta}),&\bar{\gamma}_{\eta'}= \frac{N_cC^{(0)}_{Z^0}}{2\pi^2\Delta}(f^{(3)}_{\pi^0}f^{(8)}_{\eta} - f^{(8)}_{\pi^0}f^{(3)}_{\eta}). \label{coef_etaprime_Z}
	\end{align}

As it was noted earlier in text, for the case of $\gamma\gamma^{*}$ in paper \cite{Khlebtsov:2020rte} it was shown that for ASR \eqref{asr38aa} and \eqref{mix0} can be analytically continued to time-like region, and one can obtain equations for TFFs in time-like domain \eqref{solution_generic_time}. In the case of $\gamma Z^{0}$ the analytical continuation of ASR \eqref{asr38aa_weak} and \eqref{mix0_weak} will be similar to the $\gamma\gamma^{*}$ case because they differ from each other only by the charge factor coefficients, which are constants. Thus the equations TFFs in time-like domain read:
\begin{equation}\label{solution_generic_Z_time}
	|F_{P\gamma Z^{0}}(Q^2)| = \left | \bar{\alpha}_{P} \frac{s_3}{s_3-q^2} + \bar{\beta}_{P} \frac{s_8}{s_8-q^2} + \bar{\gamma}_{P} \frac{s_0}{s_0-q^2}[1+B_{Z^0}(q^2,s_0)] \right |,
\end{equation}
where $P = \pi^0, \eta, \eta'$. 
	
Thus one can use the equations for $\pi^0,\eta,\eta'$ meson TFFs \eqref{solution_generic_Z_time} and predict decay widths of $\pi^0,\eta,\eta' \rightarrow \nu \bar{\nu} \gamma$ processes
\begin{align}\label{width}
\Gamma_{P \rightarrow \nu \bar{\nu} \gamma} & = \frac{1}{24}\frac{M_{Z^0}^4}{(M_{Z^0}^2-q^2)^2}\frac{1}{\pi^2}\frac{\alpha_{qed}G_F^2}{m_{P}^3}\int_0^{m_{P}^2}q^2(m_{P}^2-q^2)^3|F_{P\gamma Z^{0}}(q^2)|^2dq^2,
\end{align}
where $P=\pi^0,\eta,\eta'$. The decay width equation \eqref{width} is for neutrino of one flavor, in order to calculate for all 3 neutrino flavors, one should multiply \eqref{width} by a factor 3.  

In the paper \cite{Arnellos:1981bk} by Arnellos et.al. the estimates for $\Gamma_{\pi^0,\eta,\eta' \rightarrow \nu \bar{\nu} \gamma}$ are listed. These estimates were done under assumption that $q^2\leqslant m_{\pi^0,\eta}^2\approx0$. So one can calculate the same estimates with $|F_{P\gamma Z^{0}}(0)|^2$
\begin{align}\label{width_arnellos}
\Gamma_{P \rightarrow \nu \bar{\nu} \gamma} & = \frac{\alpha_{qed}G_F^2m_{P}^7}{480\pi^2}|F_{P\gamma Z^{0}}(0)|^2,
\end{align}
where $P=\pi^0,\eta,\eta'$. The results are listed below:
\begin{equation}\label{arnellos_pi0}
\Gamma_{\pi^0 \rightarrow \nu \bar{\nu} \gamma} = 1.6\cdot10^{-26} GeV,
\end{equation}
\begin{equation}\label{arnellos_eta}
\Gamma_{\eta \rightarrow \nu \bar{\nu} \gamma} = 2.6\cdot10^{-21} GeV. 
\end{equation}
Let us stress  that the value of $\Gamma_{\pi^0 \rightarrow \nu \bar{\nu} \gamma}$ \eqref{arnellos_pi0} is calculated for one term equation for $\pi^0$ TFF, i.e. without taking into account small mixing between $\pi^0$ and $\eta-\eta'$. The value of $\Gamma_{\eta \rightarrow \nu \bar{\nu} \gamma}$ \eqref{arnellos_eta} is calculated taking into account strong mixing between $\eta$ and $\eta'$, but using old mixing scheme. Also note that the results for $\pi^0$ \eqref{arnellos_pi0} and $\eta$ \eqref{arnellos_eta} are calculated for the single neutrino flavor.

It is  known that  $\eta-\eta'$ system manifests the  strong mixing, while mixing between $\pi^0$ and $\eta-\eta'$ is much smaller. However, in the paper \cite{Khlebtsov:2020rte} it was shown that even this small mixing between $\pi^0$ and $\eta-\eta'$ are needed to be taken into account in order to get proper description of $\pi^0$ TFF especially in the annihilation domain in the time-like region. Let us stress that $\pi^0-\eta-\eta'$ mixing leads to appearance of  {\it three} terms in pion TFF \eqref{solution_generic_time} which is a direct consequence of quantum field theory. 

Thus in order to compare  the results  for $\pi^0$ \eqref{arnellos_pi0} and $\eta$ \eqref{arnellos_eta} calculated by Arnellos et.al. \cite{Arnellos:1981bk} in 1982 with the current calculation in ASR approach using \eqref{width_arnellos}, we consider two cases: without taking into account small mixing between $\pi^0$ and $\eta-\eta'$, and with taking into account  this small mixing:
\begin{itemize}\label{width_values}
\item MIXING OFF: 
\begin{table}[H]
\centering
		\caption{The $\Gamma_{\pi^0,\eta \rightarrow \nu \bar{\nu} \gamma}$ decay width calculated using \eqref{width_arnellos} for different mixing schemes without taking into account small mixing between $\pi^0$ and $\eta-\eta'$. The results correspond to  the single neutrino flavor .}
		\label{table7}
\begin{tabular}{l|c|c|c|c}
 & FKS98 \cite{Feldmann:1998vh}, GeV & EF05 \cite{Escribano:2005qq}, GeV & KOT12 \cite{Klopot:2012hd}, GeV & EGMS16 \cite{Escribano:2015yup}, GeV  \\ \hline
$\Gamma_{\pi^0 \rightarrow \nu \bar{\nu} \gamma}$ & $2.05\cdot10^{-26}$ & $2.05\cdot10^{-26}$ & $2.05\cdot10^{-26}$ & $2.05\cdot10^{-26}$ \\ \hline
$\Gamma_{\eta \rightarrow \nu \bar{\nu} \gamma}$ & $2.21\cdot10^{-20}$ & $2.30\cdot10^{-20}$ & $2.08\cdot10^{-20}$ & $2.12\cdot10^{-20}$  \\ 
\end{tabular}
\end{table}	
\item MIXING ON: 
\begin{table}[H]
\centering
		\caption{The $\Gamma_{\pi^0,\eta \rightarrow \nu \bar{\nu} \gamma}$ decay width calculated using \eqref{width_arnellos} for different mixing schemes with taking into account small mixing between $\pi^0$ and $\eta-\eta'$. The results correspond to the single neutrino flavor.}
		\label{table8}
\begin{tabular}{l|c|c|c|c}
 & FKS98 \cite{Feldmann:1998vh}, GeV & EF05 \cite{Escribano:2005qq}, GeV & KOT12 \cite{Klopot:2012hd}, GeV & EGMS16 \cite{Escribano:2015yup}, GeV  \\ \hline
$\Gamma_{\pi^0 \rightarrow \nu \bar{\nu} \gamma}$ & $5.60\cdot10^{-26}$ & $5.42\cdot10^{-26}$ & $5.61\cdot10^{-26}$ & $5.55\cdot10^{-26}$ \\ \hline
$\Gamma_{\eta \rightarrow \nu \bar{\nu} \gamma}$ & $2.20\cdot10^{-20}$ & $2.30\cdot10^{-20}$ & $2.07\cdot10^{-20}$ & $2.12\cdot10^{-20}$  \\ 
\end{tabular}
\end{table}	
\end{itemize}

\begin{table}[H]
\centering
		\caption{The coefficients $\bar{\alpha_{P}},\bar{\beta_{P}},\bar{\gamma_{P}}$ \eqref{coef_pi_Z}, \eqref{coef_eta_Z}, \eqref{coef_etaprime_Z} in GeV$^{-1}$  for $\gamma Z^{0}$ processes with and without taking into account mixing between $\pi^0$ and $\eta-\eta'$.}
		\label{table5}
\begin{tabular}{l|c|c|c|c|c|c|c|c|c}
\hline
\multirow{2}{*}{Mix. sch.} & \multicolumn{3}{c|}{$\pi^0$} & \multicolumn{3}{c|}{$\eta$} & \multicolumn{3}{c}{$\eta'$} \\ \cline{2-10} 
			& $\bar{\alpha}_{\pi^0}$ & $\bar{\beta}_{\pi^0}$ & $\bar{\gamma}_{\pi^0}$ & $\bar{\alpha}_{\eta}$ & $\bar{\beta}_{\eta}$ & $\bar{\gamma}_{\eta}$ & $\bar{\alpha}_{\eta'}$ & $\bar{\beta}_{\eta'}$  & $\bar{\gamma}_{\eta'}$    \\ \hline
MIX. OFF FKS98 \cite{Feldmann:1998vh} & 0.011 & 0 & 0 & 0 & 0.005 & 0.078 & 0 & -0.0008 & 0.197 \\ 
MIX. ON  FKS98 \cite{Feldmann:1998vh} & 0.011 & -2.2e-05 & 0.007 & -6.2e-05 & 0.005 & 0.077 & -0.0003 & -0.0008 & 0.197 \\ \hline

MIX. OFF EF05 \cite{Escribano:2005qq} & 0.011 & 0 & 0 & 0 & 0.004 & 0.078 & 0 & -0.0002 & 0.184 \\ 
MIX. ON  EF05 \cite{Escribano:2005qq} & 0.011 & -1.6e-07 & 0.007 & -6.9e-05 & 0.005 & 0.078 & -0.0003 & -0.0002 & 0.184 \\ \hline

MIX. OFF KOT12 \cite{Klopot:2012hd} & 0.011 & 0 & 0 & 0 & 0.005 & 0.083 & 0 & -0.0008 & 0.220 \\ 
MIX. ON  KOT12 \cite{Klopot:2012hd} & 0.011 & -2.1e-05 & 0.008 & -6.7e-05 & 0.005 & 0.083 & -0.0003 & -0.0008 & 0.219 \\ \hline

MIX. OFF  EGMS16\cite{Escribano:2015yup} & 0.011 & 0 & 0 & 0 & 0.005 & 0.079 & 0 & -0.0006 & 0.204 \\ 
MIX. ON  EGMS16\cite{Escribano:2015yup} & 0.011 & -1.5e-05 & 0.007 & -6.4e-05 & 0.005 & 0.08 & -0.0003 & -0.0006 & 0.204 \\ 
\end{tabular}
\end{table}	

\begin{table}[H]
\centering
		\caption{The coefficients $\alpha_{P},\beta_{P},\gamma_{P}$ \eqref{coef_pi}, \eqref{coef_eta}, \eqref{coef_etaprime} in GeV$^{-1}$  for $\gamma \gamma{*}$ processes with and without taking into account mixing between $\pi^0$ and $\eta-\eta'$. }
		\label{table6}
\begin{tabular}{l|c|c|c|c|c|c|c|c|c}
\hline
\multirow{2}{*}{Mix. sch.} & \multicolumn{3}{c|}{$\pi^0$} & \multicolumn{3}{c|}{$\eta$} & \multicolumn{3}{c}{$\eta'$} \\ \cline{2-10} 
			& $\alpha_{\pi^0}$ & $\beta_{\pi^0}$ & $\gamma_{\pi^0}$ & $\alpha_{\eta}$ & $\beta_{\eta}$ & $\gamma_{\eta}$ & $\alpha_{\eta'}$ & $\beta_{\eta'}$  & $\gamma_{\eta'}$    \\ \hline
MIX. OFF FKS98 \cite{Feldmann:1998vh} & 0.274 & 0 & 0 & 0 & 0.127 & 0.144 & 0 & -0.021 & 0.365 \\ 
MIX. ON  FKS98 \cite{Feldmann:1998vh} & 0.274 & -0.0005 & 0.013 & -0.0015 & 0.127 & 0.144 & -0.008 & -0.021 & 0.365 \\ \hline

MIX. OFF EF05 \cite{Escribano:2005qq} & 0.274 & 0 & 0 & 0 & 0.112 & 0.144 & 0 & -0.005 & 0.341 \\ 
MIX. ON  EF05 \cite{Escribano:2005qq} & 0.274 & 3.9e-06 & 0.012 & -0.0017 & 0.112 & 0.145 & -0.007 & -0.005 & 0.341 \\ \hline

MIX. OFF KOT12 \cite{Klopot:2012hd} & 0.274 & 0 & 0 & 0 & 0.135 & 0.154 & 0 & -0.021 & 0.407 \\ 
MIX. ON  KOT12 \cite{Klopot:2012hd} & 0.274 & -0.0005 & 0.014 & -0.0017 & 0.134 & 0.154 & -0.009 & -0.021 & 0.406 \\ \hline

MIX. OFF  EGMS16\cite{Escribano:2015yup} & 0.274 & 0 & 0 & 0 & 0.128 & 0.147 & 0 & -0.016 & 0.377 \\ 
MIX. ON  EGMS16\cite{Escribano:2015yup} & 0.274 & -0.0004 & 0.013 & -0.0016 & 0.128 & 0.147 & -0.008 & -0.016 & 0.377 \\ 
\end{tabular}
\end{table}	
Firstly, let us discuss results for $\pi^0$ meson. As it is seen in the Table \ref{table7}, when small mixing of $\pi^0$ with $\eta-\eta'$  is neglected, the decay widths calculated by ASR approach and by Arnellos et.al. \cite{Arnellos:1981bk} coincide. However when mixing between $\pi^0$ and $\eta-\eta'$ is taken into account, we got 3 times increase of decay width $\Gamma_{\pi^0 \rightarrow \nu \bar{\nu} \gamma}$ by ASR as listed in the Table \ref{table8}. This effect is an essential feature of small non-zero mixing between $\pi^0$ and $\eta-\eta'$.  As it is seen in the Table. \ref{table5}, the coefficient $\bar{\gamma}_{\pi^0}$ appears to be comparable with $\bar{\alpha}_{\pi^0}$, and so it leads to 3 times increasing of $|F_{P\gamma Z^{0}}(0)|^2$ and corresponding decay width. Note that as it can be seen from the Table \ref{table5} it wouldn't be such increasing for $\gamma\gamma^{*}$ processes, where coefficient $\alpha_{\pi^0}$ dominates. But in the case $\gamma Z^0$ the Salam-Weinberg $(\frac{1}{2}-2\xi)$ coefficient leads to lowering of $\bar{\alpha}_{\pi^0}$ in such a way that it becomes comparable to $\bar{\gamma}_{\pi^0}$. Thus for $\gamma Z^0$ case the small mixing between $\pi^0$ and $\eta-\eta'$ becomes crucial! Let us  also recall, as we pointed out earlier, that $\pi^0$ TFF must have 3 terms, which can be achieved only if mixing is not neglected.

Now let us consider the results for $\eta$ meson. As it is seen in the Table \ref{table5}, taking into account the effects of small mixing of $\pi^0$ with $\eta-\eta'$ does not have such a crucial role in contrast to $\pi^0$ case, and the leading coefficients of $\eta$ meson TFF --  $\bar{\beta}_{\eta}$ and $\bar{\gamma}_{\eta}$ do not change significantly. The coefficient $\bar{\alpha}_{\eta}$ appears to be much smaller than $\bar{\beta}_{\eta}$ and $\bar{\gamma}_{\eta}$, and so it has a small influence on $\eta$ meson TFF and correspondingly to decay width. 

At the same time, as it is seen in the Tables \ref{table7} and \ref{table8}, the obtained results for $\eta$ meson decay widths \eqref{width_values} differ from the ones estimated by Arnellos et.al. \cite{Arnellos:1981bk} by an  order of magnitude for all considered mixing schemes. This effect is a consequence of the choice of the mixing scheme. In the paper \cite{Arnellos:1981bk}, which was written in 1982, the old mixing scheme  was used and so the values of decay constants differ from the modern ones. The use of recent mixing schemes as it is seen in the Tables \ref{table7} and \ref{table8} leads to the growth of the $\Gamma_{\eta \rightarrow \nu \bar{\nu} \gamma}$ decay width by an order of magnitude.

Let us point out that, strictly speaking, for the case of $\eta$ the interval of $dq^2\leqslant m_{\eta}^2\approx0.3$ GeV$^2$ is much larger than in the $\pi^0$ case $dq^2\leqslant m_{\pi^0}^2\approx0.02$ GeV$^2$ and so the estimation of $\Gamma_{\eta \rightarrow \nu \bar{\nu} \gamma}$ by \eqref{width_arnellos} may not be accurate. Thus we need to numerically integrate \eqref{width} for the $\eta$ meson. Using instruments of ROOT \cite{root} numerical integration of \eqref{width} for the EGMS16\cite{Escribano:2015yup} mixing scheme with interpolation formulas for $s_{3,8,0}(q^2)$ \eqref{s8_new},\eqref{s0_new} and $B(q^2)$ \eqref{b_new} with the values of the corresponding coefficients listed in the Table.\ref{table3} gives 
\begin{equation}\label{width_one_neutrino}
\Gamma_{\eta \rightarrow \nu \bar{\nu} \gamma} = 3.45\cdot10^{-20} \ GeV,
\end{equation}
for each  neutrino flavor. One can see that taking into account non-trivial form of $\eta$ TFF leads to increasing of the $\Gamma_{\eta \rightarrow \nu \bar{\nu} \gamma}$ by the factor 1.5 in comparison of the value calculated using \eqref{width_arnellos}. In order to obtain result for all 3 neutrino flavors one should multiply \eqref{width_one_neutrino} by a factor 3, so finally
\begin{equation}\label{width_3_neutrino}
\Gamma_{\eta \rightarrow \sum_i(\nu_i \bar{\nu_i}) \gamma} = 1.04\cdot10^{-19} \ GeV.
\end{equation}
Despite the obtained result is by 2 orders of magnitude higher than 1982 year calculation \eqref{arnellos_eta} by Arnellos et.al. \cite{Arnellos:1981bk} it is still experimentally unobservable at the current level of accuracy and does not therefore constitute the substantial    
background for dark matter searches in the invisible mode.

For  $\pi^0$ one can also perform numerical integration. The result for the EGMS16\cite{Escribano:2015yup} mixing scheme with interpolation formulas for $s_{3,8,0}(q^2)$ \eqref{s8_new},\eqref{s0_new} and $B(q^2)$ \eqref{b_new} with the values of the corresponding coefficients listed in the Table.\ref{table3} reads 
\begin{equation}\label{width_one_neutrino_pion}
\Gamma_{\pi^0 \rightarrow \nu \bar{\nu} \gamma} = 5.68\cdot10^{-26} \ GeV,
\end{equation}
for  each of neutrino flavor . And for all 3 flavors
\begin{equation}\label{width_3_neutrino_pion}
\Gamma_{\pi^0 \rightarrow \sum_i(\nu_i \bar{\nu_i}) \gamma} = 1.704\cdot10^{-25} \ GeV.
\end{equation}

The modern experimental upper limit for $\Gamma_{\pi^0 \rightarrow \sum_i(\nu_i \bar{\nu_i}) \gamma}$ is $< 1.48\cdot10^{-17}$ GeV by NA62 collaboration \cite{NA62:2019meo}.

\textbf{The branching for $\pi^0$ and $\eta$ meson are the following:}
\begin{equation}\label{branching_pion}
\frac{\Gamma_{\pi^0 \rightarrow \sum_i(\nu_i \bar{\nu_i}) \gamma}}{\Gamma_{\pi^0 \rightarrow all}} = 2.18\cdot10^{-17}, 
\end{equation}

\begin{equation}\label{branching_eta}
\frac{\Gamma_{\eta \rightarrow \sum_i(\nu_i \bar{\nu_i}) \gamma}}{\Gamma_{\eta \rightarrow all}} = 7.94\cdot10^{-14}.
\end{equation}

In order to avoid possible uncertainties in determination of $\pi^0$ and $\eta$ TFFs \eqref{solution_generic_Z} connected with the choice of the mixing scheme and higher order corrections we propose to consider the ratio between $\frac{d\Gamma_{\eta \rightarrow \sum_i(\nu_i \bar{\nu_i}) \gamma}}{dq^2}$ and $\Gamma_{\pi^0 \rightarrow \sum_i(\nu_i \bar{\nu_i}) \gamma}$. The graph of this ratio shown in the Fig.\ref{fig:201} for the EGMS16\cite{Escribano:2015yup} mixing scheme with interpolation formulas for $s_{3,8,0}(q^2)$ \eqref{s8_new},\eqref{s0_new} and $B(q^2)$ \eqref{b_new} with the values of the corresponding coefficients listed in the Table.\ref{table3}. The $\Gamma_{\pi^0 \rightarrow \nu \bar{\nu} \gamma}$ decay width is taken from \eqref{width_3_neutrino_pion}. 
\begin{figure}[H]
\caption{The $\frac{d\Gamma_{\eta \rightarrow \sum_i(\nu_i \bar{\nu_i}) \gamma}}{dq^2}$ graph normalised on $\Gamma_{\pi^0 \rightarrow \sum_i(\nu_i \bar{\nu_i}) \gamma}$ for the EGMS16\cite{Escribano:2015yup} mixing scheme with interpolation formulas for $s_{3,8,0}(q^2)$ \eqref{s8_new},\eqref{s0_new} and $B(q^2)$ \eqref{b_new} with the values of the corresponding coefficients listed in the Table.\ref{table3}. The $\Gamma_{\pi^0 \rightarrow \nu \bar{\nu} \gamma}$ decay width taken from \eqref{width_3_neutrino_pion}. }
\includegraphics[width=1.0\textwidth]{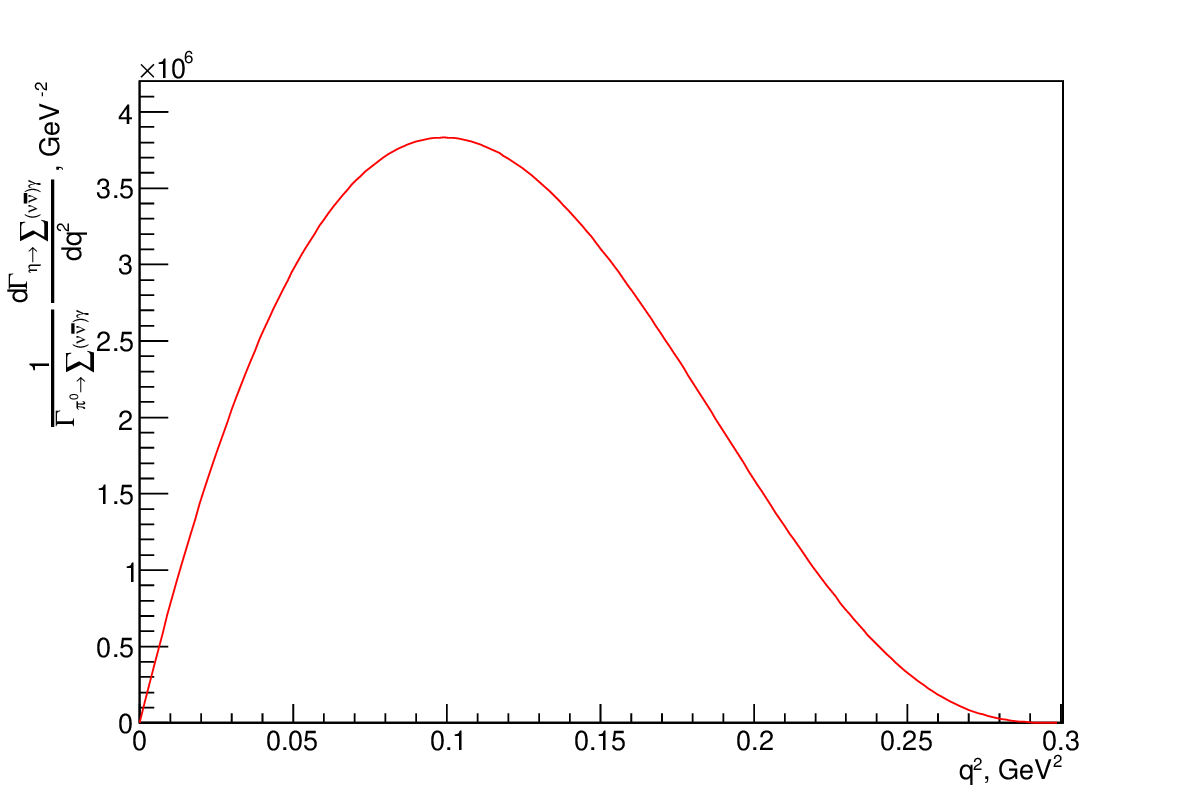}\label{fig:201}
\end{figure}

\section{Discussion and Conclusions}

In the present paper we addressed the various decays of pseudoscalar mesons.  The analysis included both "visible" (to
two photons or to a photon and charged lepton pair) and 
"invisible" or "semi-visible" when instead of charged leptons pair the neutrino-antineutrino pair is produced.

Their consideration appeared to be mutually dependent. Namely, the analysis of the available data for visible mode 
allowed us to extract the important information to describe the
invisible one.      

To achieve this goal the dispersive approach to the axial anomal was used, both Abelian and non-Abelian cases are considered. Anomalous sum rules for isovector, octet and singlet currents are derived, which allow us to extract the nonperturbative gluon matrix element $\langle 0 |G\tilde{G}|\gamma\gamma^{(*)} \rangle$ from experimentally measured transition form factors $\pi^0,\eta,\eta'\rightarrow\gamma\gamma^{(*)}$ and describe it as a function of the photon virtuality. 

For analysis of invisible mode it is important that 
this matrix element differs from the matrix element $\langle 0 |G\tilde{G}|\gamma Z^{0} \rangle$ up to $\frac{1}{m_{W}^2}$ corrections only by a charge factor. Thus one can predict behaviour of the matrix element $\langle 0 |G\tilde{G}|\gamma Z^{0} \rangle$ as a function of $Z^0$ momentum in $\pi^{0},\eta,\eta'\rightarrow \nu \bar{\nu} \gamma$ Dalitz decays. The decay widths of such processes were calculated:}
$$\Gamma_{\pi^0 \rightarrow \sum_i(\nu_i \bar{\nu_i}) \gamma} = 1.704\cdot10^{-25} GeV,$$
$$\Gamma_{\eta \rightarrow \sum_i(\nu_i \bar{\nu_i}) \gamma} = 1.04\cdot10^{-19} GeV.$$

Our result for pion decay coincides with that of Arnellos, Marciano and Parsa only if its small mixing is neglected. 
As at such approximation the result is proportional to the small factor $4 sin^2 \theta_W - 1$, account for this mixing leads to the increase of the result by factor 3. In turn, due to use of modern mixing schemes, the $\eta$ decay branching is enhanced by two orders of magnitude.

Despite this enhancement, invisible $\eta$ decay, which may be a background for the decay to dark photon, is not playing this role at the current level of accuracy. 

At the same time, the accuracy may be substantially increased
in the experiments at HHaS detector at HIAF facility. 
Moreover, the increased accuracy of investigations of  visible Dalitz decay may provide more information on the non-perturbative gluon form factor $B(q^2)$ and investigate in detail its non-trivial behavior at $q^2 \sim 1 GeV^2$.

This beaviour may be related to the existence of pseudoscalar 
glueball with a mass about 1.5-2 GeV \cite{Khlebtsov:2020rte}.
Taking into account the necessity of phase space to provide the Dalitz glueball decay related to  $B(q^2)$,  one may even think about $X(2370)$
in such a role.

Another possible role of  $B(q^2)$ corresponds to the application of TFF for description of decay to dark photon,
where  $B(m_{Dark}^2)$ should enter instead. The mentioned structure in  $B(q^2)$ may be used to contribute to the current searches of dark photon in this region    
\cite{Gorbunov:2023jnx,Gorbunov:2024vrc,Gorbunov:2025sth}. In the case of the dark photon mass around 1 GeV even the extremum of  $B(q^2)$ can play a role in the case of invisible decays of heavy pseudoscalars, like the mentioned glueball. These effects deserve further investigation. 

We are indebted to N.V. Krasnikov and A.S. Zhevlakov for useful comments.

 O.T. is grateful to Southern Center for Nuclear Theory  of the Institute of Modern Physics CAS, where the paper was started and completed, for warm hospitality.
His stays were supported by the CAS
President’s International Fellowship Initiative.

\section*{Appendix A}\label{appendix_a}
We list here  the values of decay constants used in the present work and which were evaluated in several other analyses: the decay constants of the $\eta-\eta'$ mixing ($f^{(8,0)}_{\eta,\eta'}$) were taken from Refs. \cite{Feldmann:1998vh,Escribano:2005qq,Klopot:2012hd,Escribano:2015yup}, and the constants of the $\pi^0$ admixtures to the $\eta-\eta'$ system ($f^{(3)}_{\eta,\eta'}$ and $f_{\pi}^{(8,0)}$) were taken from \cite{Escribano:2020jdy}. The mixing parameters evaluated in the cited works were expressed in terms of the decay constants.  The pion decay constant is $f_{\pi^0}^{(3)}=f_{\pi}=0.1307$~GeV.

\begin{equation} \label{fks98}
	\textbf{FKS98}\cite{Feldmann:1998vh}:\left(
	\begin{matrix}
		f_{\pi^0}^{(3)} & f_{\eta}^{(3)} & f_{\eta'}^{(3)} \\
		f_{\pi^0}^{(8)} & f_{\eta}^{(8)} & f_{\eta'}^{(8)} \\
		f_{\pi^0}^{(0)} & f_{\eta}^{(0)} & f_{\eta'}^{(0)} 
	\end{matrix}
	\right)=\left(
	\begin{matrix}
		1 & -0.0015 & -0.035 \\
		-0.0066 & 1.17 & -0.46 \\
		0.034 & 0.19 & 1.15 
	\end{matrix}
	\right) f_{\pi},
\end{equation}

\begin{equation} \label{EF05}
	\textbf{EF05}\cite{Escribano:2005qq}:\left(
	\begin{matrix}
		f_{\pi^0}^{(3)} & f_{\eta}^{(3)} & f_{\eta'}^{(3)} \\
		f_{\pi^0}^{(8)} & f_{\eta}^{(8)} & f_{\eta'}^{(8)} \\
		f_{\pi^0}^{(0)} & f_{\eta}^{(0)} & f_{\eta'}^{(0)} 
	\end{matrix}
	\right)=\left(
	\begin{matrix}
		1 & -0.0015 & -0.035 \\
		-0.0066 & 1.39 & -0.59 \\
		0.034 & 0.054 & 1.29 
	\end{matrix}
	\right) f_{\pi},
\end{equation}

\begin{equation} \label{KOT12}
	\textbf{KOT12}\cite{Klopot:2012hd}:\left(
	\begin{matrix}
		f_{\pi^0}^{(3)} & f_{\eta}^{(3)} & f_{\eta'}^{(3)} \\
		f_{\pi^0}^{(8)} & f_{\eta}^{(8)} & f_{\eta'}^{(8)} \\
		f_{\pi^0}^{(0)} & f_{\eta}^{(0)} & f_{\eta'}^{(0)} 
	\end{matrix}
	\right)=\left(
	\begin{matrix}
		1 & -0.0015 & -0.035 \\
		-0.0066 & 1.11 & -0.42 \\
		0.034 & 0.16 & 1.04 
	\end{matrix}
	\right) f_{\pi},
\end{equation}

\begin{equation} \label{EGMS16}
	\textbf{EGMS16}\cite{Escribano:2015yup}:\left(
	\begin{matrix}
		f_{\pi^0}^{(3)} & f_{\eta}^{(3)} & f_{\eta'}^{(3)} \\
		f_{\pi^0}^{(8)} & f_{\eta}^{(8)} & f_{\eta'}^{(8)} \\
		f_{\pi^0}^{(0)} & f_{\eta}^{(0)} & f_{\eta'}^{(0)} 
	\end{matrix}
	\right)=\left(
	\begin{matrix}
		1 & -0.0015 & -0.035 \\
		-0.0066 & 1.18 & -0.46 \\
		0.034 & 0.14 & 1.13 
	\end{matrix}
	\right) f_{\pi},
\end{equation}

\end{document}